# Exploring triad-rich substructures by graph-theoretic characterizations in complex networks

**Songwei Jia[1], Lin Gao[1] \*, Yong Gao[2], James Nastos[2], Xiao Wen[1], Xindong Zhang[1] and Haiyang Wang[1]**

[1] School of Computer Science and Technology, Xidian University, Xi'an 710071, China

[2] Department of Computer Science, University of British Columbia Okanagan, Kelowna, British Columbia, Canada V1V 1V7

E-mail: lgao@mail.xidian.edu.cn.

**Abstract**

One of the most important problems in complex networks is how to detect metadata groups accurately. The main challenge lies in the fact that traditional structural communities do not always capture the intrinsic features of metadata groups. Motivated by the observation that metadata groups in PPI networks tend to consist of an abundance of interacting triad motifs, we define a 2-club substructure with diameter 2 which possessing triad-rich property to describe a metadata group. Based on the triad-rich substructure, we design a <u>DIV</u>ision <u>A</u>lgorithm using our proposed edge <u>N</u>iche <u>C</u>entrality DIVANC to detect metadata groups effectively in complex networks. We also extend DIVANC to detect overlapping metadata groups by proposing a simple 2-hop overlapping strategy. To verify the effectiveness of triad-rich substructures, we compare DIVANC with existing algorithms on PPI networks, LFR synthetic networks and football networks. The experimental results show that DIVANC outperforms most other algorithms significantly and, in particular, can detect sparse metadata groups.

## 1. Introduction

One of the most important problems in complex networks is how to detect metadata groups accurately [1]. Metadata groups are the subsets of vertices with real physical sense. For example, in biological networks they are referred to as various biological functional modules such as protein complexes, GO terms and pathways; in social networks, metadata groups may be various social circles such as groups of people with common interests, etc. Traditional structural communities, which are typically described as dense subgraphs (subnetworks) explicitly or implicitly, are usually used to capture the intuition of metadata groups. The underlying assumption is that objects in some metadata groups really tend to interact more frequently than in other regions of the network. Around the issue of how to detect metadata groups, scholars have proposed many popular structural community detection algorithms which can identify parts of metadata groups successfully at a certain degree. Examples of algorithms that detect metadata groups by dense subnetworks include (i) random-walk based methods such as MCL [2] and INFOMAP [3]; (ii) seed-growing methods such as MCODE [4] and ClusterOne [5]; (iii) algorithms based on clustering, optimization, or statistical techniques such as LinkComm [6], LOUVAIN [7], and OSLOM [8]; and (iv) algorithms based on deeper graph-theoretic features such as EPCA [9–11].

While detecting traditional structural communities can offer some insight into some of the structure of metadata groups, more and more recent studies show that these intuitions about traditional structural communities are unreliable [12–16]. Some of these examples include: perhaps most profoundly, metadata groups do not necessarily coincide with traditional structural communities [13–16]; overlapping communities have a higher density of links in the overlapping parts than in the non-overlapping ones, which are in contrast with the common picture of traditional structural communities [16]; there is a paradox that the detection of well-defined communities is more difficult than the identification of ill-defined communities [12]. All of these counterintuitive evidences hint at the necessity of modifying the general defining characteristics of traditional structural communities. While there is a general consensus on the fact that there is a need for an adjustment of the notion of community or clusters, there is no clear direction to a remedy. Scholars [15] point out that there are two possible scenarios for filling the gaps between traditional structural communities and metadata groups. One is to include additional topological features in refining the definitions of traditional structural communities beyond the standard measures of link density, degree correlations or density of loops, etc.; the other is to add requirements based non-topological knowledge, such as domain-specific background knowledge [17–19] for the detection of metadata groups. However, in the former case, solely adjusting the structural conditions sought for may still not obtain satisfying results as the essence of metadata groups in all contexts may not be characterized by equivalent topology. In the latter case, adding various domain-specific background knowledge may be effective on a limited number of cases, but the reliance on rigid domain-specific knowledge makes the resulting algorithms unlikely to exhibit scalability or transferability to other domains. An ideal paradigm for fully analyzing metadata groups would include identification of metadata groups via certain specific intrinsic features, combined with a method for capturing deeper domain-specific structure in a general topological framework on which one can further develop algorithms. Here, we develop such a new framework by incorporating a novel and more subtle assumption based on graph-theoretic properties of metadata groups and design efficient computing procedures to detect non-overlapping and overlapping substructures that have the desired properties. As shown in figures 1(a) and 1(b), both of the metadata groups 'nuclear origin of replication recognition complex' [20] (dense) and 'GID complex' [21] (sparse) consist of abundantly interacting triad motifs, for instance. More details about the two complexes can be found in Appendix A. Motivated by the observation that metadata groups in a PPI network are either quite dense or quite sparse and tend to consist of an abundance of interacting triad motifs [22–25], we define a 2-club substructure with diameter 2 which possessing triad-rich property to describe a metadata group. Based on the triad-rich substructure, we design a <u>DIV</u>ision <u>A</u>lgorithm using our proposed edge <u>N</u>iche <u>C</u>entrality DIVANC to detect metadata groups effectively in complex networks. We also extend DIVANC to detect overlapping metadata groups by proposing a simple 2-hop overlapping strategy. To verify the effectiveness of triad-rich substructures, we compare DIVANC with existing algorithms on PPI networks, LFR synthetic networks and football networks. The experimental results show that DIVANC outperforms most other algorithms significantly and, in particular, can detect sparse metadata groups.

The rest of the paper is organized as follows. In Section 2, we present our framework for detecting candidate metadata groups. After discussing our datasets and providing statistical evidence that motivates and supports our triad-rich assumption about metadata groups in Sections 2.1 and 2.2, we give a formal definition of a 2-club substructure in Section 2.3. In Section 2.4, we discuss the details of our algorithm for 2-club substructure detection, including a new edge-centrality measure specifically designed for 2-club substructures as well as a 2-hop-based strategy for extracting overlapping 2-club substructures. In Section 3, we report and discuss our experimental results. In Section 4, we conclude the paper and give closing discussion.

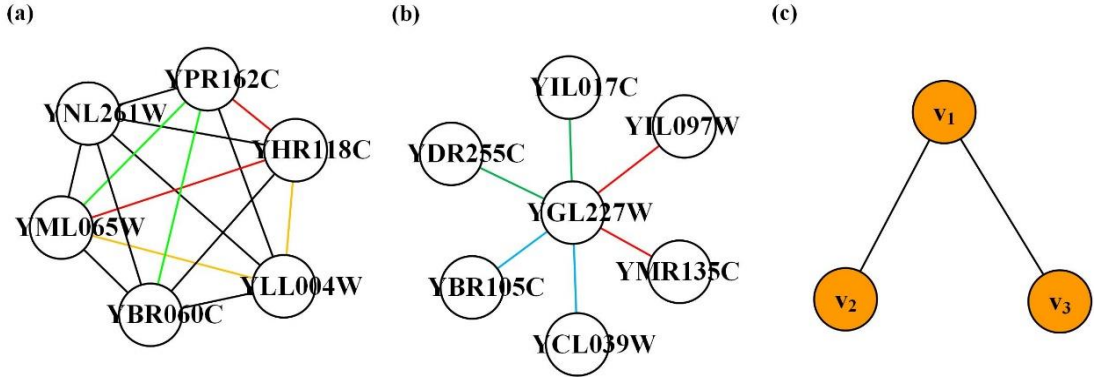

**Figure 1.** Metadata groups consisting of abundantly interacted triad motifs. (a) Nuclear origin of replication recognition complex; (b) GID complex; (c) an example of triad motif.

## 2. Methods

*2.1. The datasets*

We apply our framework on PPI networks [26,27], LFR synthetic networks [28,29] and football networks [30,31]. In the following, we give details about the relative networks and six golden standard sets of metadata groups in PPI networks, respectively.

*S.cerevisiae* PPI networks (*Sce*DIP) are obtained from DIP [26] and *H.sapiens* PPI networks (*Hsa*HPRD) are extracted from HPRD [27]. For *Sce*DIP, we use the sets from the Munich Information Center for Protein Sequences (MIPS) [32], Saccharomyces Genome Database (SGD) [33] and *S.cerevisiae* GO terms (*Sce* GO term) as golden standards [34,35]. For *Hsa*HPRD, we use the sets of Human Protein Complex Database with a Complex Quality Index (PCDq) [36], Comprehensive Resource of Mammalian Protein Complexes (CORUM) [37] and *H.sapiens* GO terms (*Hsa* GO term) [34,35] as golden standards. *Sce*DIP consists of 4980 proteins and 22076 interactions; *Hsa*HPRD consists of 9269 proteins and 36917 interactions. The GO terms are not composed of all the terms but the high-level GO terms whose information content is more than 2 [34,35]. The definition of the information content ($IC$) of a GO term $g$ is $IC = -\log(|g|/|root|)$ as given in the literature [34], where *'root'* is the corresponding root GO terms across the three aspects of molecular function (MF), biological process (BP) or cellular component (CC)) of $g$. In addition, the GO terms with less than 2 proteins are removed. We also remove the protein complexes or GO terms of which no members appear in the corresponding PPI networks. Last, MIPS consists of 203 and SGD has 305 protein complexes, while PCDq includes 1204 and CORUM has 1294 complexes. Additionally, there are 1050 terms in *Sce* GO term, and 4457 terms in *Hsa* GO term.

Here we also give the details of LFR synthetic networks [28,29]. The parameters of the series of LFR networks are: vertices size $N = 1000$, average degree $\bar{k} = 15$, minimum community size $minc = 20$, maximum community size $maxc = 50$, the mixing parameter $mu$ with a step of 0.1 from 0.1 to 1.0, and for overlapping LFR networks, additional parameters such as number of overlapping vertices $on = 100$, number of memberships of the overlapping vertices $om = 2$. The parameters were chosen to follow the examples provided by the original code and we downloaded it at http://santo.fortunato.googlepages.com/inthepress2.

Football network [30,31] represents the relationships played among college teams during the year 2001 football season of the USA, and consists of 115 vertices and 613 edges, indicating 115 teams and 613 games played against each other. The 115 teams are grouped into 11 conferences, with a 12th group of independent teams.

*2.2. Triad-rich substructures: a novel assumption on metadata groups*

As mentioned in Section 1, most existing community detection algorithms are (explicitly or implicitly) based on the assumption that metadata groups most likely appear in dense subnetworks. This edge-rich assumption results in two fundamental difficulties that make it hard, if not impossible, to improve the performance of those methods that detect metadata groups by extracting dense subnetworks: (1) the requirement for a subnetwork to be highly dense is too strong; and (2) a pure density-based measure cannot distinguish among subnetworks that have different internal structures that may be of physical significance.

To further evidence the rough assumption that metadata groups most likely appear in dense subnetworks is not comprehensive enough, we quantitatively analyze the density distributions of the metadata groups among six golden standard sets in corresponding PPI networks, for instance. The numbers of elements in golden standard sets of PPI networks are described in table 1. As shown in figure 2, we demonstrate the percentages of metadata groups among their whole golden standard sets of SGD, MIPS, *Sce* GO term, PCDq, CORUM and *Hsa* GO term according to their density distributions, respectively. We consider the percentages of metadata groups with their densities 0, greater than 0 but no more than 0.1, greater than 0.1 but no more than 0.2, …, densities 1 but sizes greater than 2, densities 1 but sizes 2 respectively, as described in the legend of figure 2. Here, we demonstrate the percentages of those with densities 1 but sizes 2 since although these metadata groups seem very dense, they are merely paths with two vertices. As shown in figure 2, there are only a small number of metadata groups with high density in the golden standard sets of SGD, MIPS, *Sce* GO term, PCDq, CORUM and *Hsa* GO term.

Motivated by the observation that metadata groups (either dense or sparse) in a PPI tend to consist of abundantly interacting triad motifs [22–25], we propose that a metadata groups detection method shall be based on the following triad-rich assumption: metadata groups are most likely to occur in the substructures that contain many interacting triads. A motif in complex networks is a pattern of subnetworks on a small number of vertices that occur at a significantly higher frequency than what is expected in a random network with similar network statistics. In this paper, we will focus on the most basic blocks of triad motif [38] that consist of 3-vertices and two links, as depicted in figure 1(c). This is because (i) a 2-vertices motif is nothing but an edge, and is trivial; and (ii) motifs containing more than three vertices can be constructed from interacting triad motifs.

The triad-rich assumption naturally generalizes the edge-rich assumption in that a dense subnetwork is triad-rich. For example, the 'Nuclear origin of replication recognition complex' shown in figure 1(a) is a clique with the largest density, it is not hard to see that the complex contains many interacting triad motifs. However, a triad-rich subnetwork is not necessarily edge-rich, making it possible to detect metadata groups that are not necessarily dense. For example, the 'GID complex' shown in figure 1(b) is a star subnetwork with the lowest density but containing many interacting triad motifs. To quantify the property of being triad-rich, we impose the requirement that every pair of vertices in a metadata group participates in at least one triadic interaction. This leads to the graph-theoretic definition of a triad-rich metadata group as a substructure with diameter 2 and triad-rich property (i.e., a 2-club). This definition of a triad-rich substructure makes it possible to study interesting internal structures of metadata groups, which cannot be distinguished by any density-mainly measure.

To further support the proposed triad-rich assumption that metadata groups are more likely to occur in substructures that contain many interacting triads, we give a basic statistical analysis on the distribution of triad motifs in metadata groups on PPI networks. For PPI networks, we use the above six golden standard sets described in Section 2.1 as metadata groups, we compare the frequencies of triad motifs in metadata groups to those of a random selection of equally-sized subnetworks. Our process is as follows: we first count the number of triads existing in metadata groups. For each of the metadata groups, if it contains $n$ vertices, we randomly choose a set of $n$ vertices in the corresponding PPI network and count the triad motifs among those randomly-selected vertices. We repeat this random selection one thousand times and calculate the corresponding average triad number. Thus for each golden standards, we obtain a pair of vectors, one of which indicates the triad numbers for the metadata groups and the other indicating the average triad numbers for the subnetworks obtained by randomly choosing vertices. The dimensions of the vectors are equal to the numbers of metadata groups in their golden standard sets. For instance, for the MIPS, we have a vector of 203 values of the true counts of triad motifs in the 203 complexes, along with a vector of triad counts in randomly generated subnetworks of equal size to each of the complexes. To test the statistical significance for the triad distributions among each golden standard sets, we calculate the corresponding P-values based on a T-test by comparing the number of triads obtained in the metadata groups to the numbers obtained in the randomly-selected equal-sized subnetworks. The lower P-values mean the more significant triad distribution in metadata groups. We display the corresponding P-values for MIPS, SGD, PCDq, CORUM, *Sce* GO term and *Hsa* GO term in table 1 respectively, where it is readily seen that the randomly-selected subnetworks have statistically fewer triad motifs than the true benchmarks. Thus, triad motifs are distributed far more densely in metadata groups than in randomly-selected subnetworks. This result reinforces our proposed novel assumption that metadata groups consist of abundantly interacted triad motifs.

**Table 1.** The details of six golden standard sets and their corresponding P-values of triad distribution in PPI networks.

| Golden standards | MIPS | SGD | PCDq | CORUM | *Sce* GO term | *Hsa* GO term |
|---|---|---|---|---|---|---|
| Numbers | 203 | 305 | 1204 | 1294 | 1050 | 4457 |
| P-value | 6.98e-08 | 3.69e-05 | 2.49e-15 | 2.46e-18 | 4.79e-11 | 1.97e-52 |

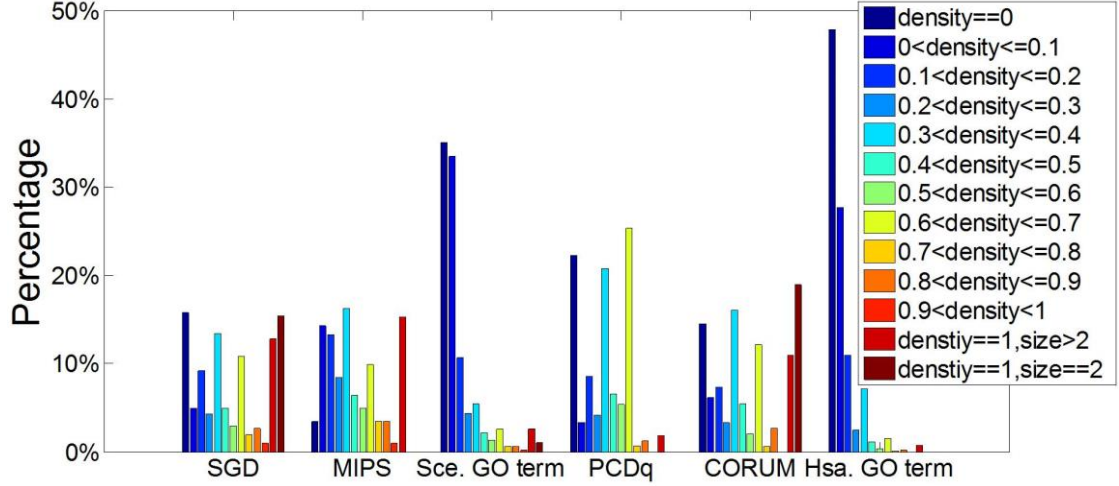

**Figure 2.** Density distribution among six golden standard sets in PPI networks.

*2.3. A graph-theoretic definition of triad-rich substructures*

In this section, we mainly introduce relative terminologies and our graph-theoretic definition of a triad-rich substructure.

*2.3.1. Terminologies and concepts in graph theory.* A graph or network $G = (V, E)$ consists of a vertex set $V$ and an edge set $E$. An induced subgraph of a graph is specified by a set of vertices, and all of the edges that exist on those vertices in the network are also part of the induced subgraph [39]. A $P_4$ is an induced graph on four ordered vertices, which are connected as a simple path [11,39]. The distance between two vertices is the length (i.e., the number of edges) of a shortest path between them. The diameter of a graph is the maximum distance between a pair of vertices.

*2.3.2. 2-club substructures.* We define a triad-rich substructure to be an induced and connected substructure where every pair of vertices participate in at least one triadic interaction. It is not hard to see that a substructure is a triad-rich one if and only if it is an induced subgraph of diameter 2. Noting that a diameter-2 induced subgraph is also known as a 2-club in the social network literature, we shall call our triad-rich substructure a 2-club substructure. In a 2-club substructure, every pair of vertices either form an edge or are contained in at least one triad motif and in fact, 2-club substructures play the same role in the class of triad-rich subnetworks as cliques do in the class of edge-rich subnetworks.

*2.4. DIVANC: a division algorithm for finding 2-club substructures*

Given the definition of a triad-rich substructure as a 2-club, a natural algorithmic problem is to delete the minimum number of edges to modify a network into a new network where each connected component has diameter 2. While we have not been able to give a formal proof, we believe that this problem is NP-hard, similar to the many NP-hard edge-deletion problems. In this section, we develop a new centrality measure to approximate the requirement of being of diameter 2, and design effective and efficient algorithm for detecting both non-overlapping and overlapping 2-club substructures. Our algorithm is an edge division algorithm that removes edges according to a new edge centrality measure, called the edge niche centrality, specifically designed to capture the properties of 2-club substructures.

*2.4.1. Edge niche centrality.* In their seminal work, Girvan and Newman [30] proposed an edge-division algorithm to detect communities by iteratively removing edges with high edge betweenness centrality. One of the issues with the G-N algorithm is the high time complexity of computing the edge betweenness centrality, even though there are polynomial time algorithms for it. Recently, a few easy-to-compute centrality measures have been proposed and used to design more efficient edge-division algorithms, including the $P_4$ centrality [11], anti-triangle centrality [40], and the edge clustering coefficient [41].

The new edge-centrality measure, edge niche centrality, measures the importance of an edge by taking into consideration the edge's $P_4$ centrality and embeddedness (revealed by edge clustering coefficient). In the following, we give the formal definition of our edge niche centrality. Let $G = (V, E)$ be an undirected and unweighted network and $e_{ij}$ be an arbitrary edge in $G$, the niche centrality $C_{ij}^N$ of $e_{ij}$ is defined as:

$$C_{ij}^N = C_{ij}^- + \min(k_i - 1, k_j - 1) / (C_{ij}^\Delta + 1) \qquad (1)$$

where $C_{ij}^-$ represents the edge $P_4$ centrality defined as the number of $P_4$s which $e_{ij}$ belongs to, and can be calculated by the function $IsP_4(a, b, c, d)$ provided in the reference [11] simply; $C_{ij}^\Delta$ is the number of triangles which $e_{ij}$ belongs to, representing the embeddedness of $e_{ij}$ (i.e., the number of common neighbors of vertices $v_i$ and $v_j$); $k_i$ ($k_j$) denotes the degree of the vertex $v_i$ ($v_j$).

As shown in Equation (1), two factors are considered in the edge niche centrality. The $P_4$ centrality, helps in identifying edges that participate in many induced paths of length 3. Removing edges with high $P_4$ centrality helps separating vertices that have distance greater than 2 and therefore, are not likely to be in the same 2-club substructure. The second term distinguishes edges that have similar $P_4$ centrality, but have different embeddedness. This definition of edge niche centrality gives us a way to quantitatively measure the extent to which an edge is an inter-link or intra-link. If its niche centrality is large the edge is more likely to be an inter-link, while if its niche centrality is smaller it is more likely to be an intra-link.

*2.4.2. The 2-hop overlapping strategy.* As edge-division algorithms can only detect non-overlapping substructures, we propose a strategy, the 2-hop overlapping strategy, to uncover 2-club substructures that may overlap. Our strategy is inspired by the idea of overlapping communities in [42]. It searches eligible peripheral vertices and adds them into non-overlapping substructures to obtain the corresponding overlapping 2-club substructures. The criterion used to add a peripheral vertex is based on its closeness to a 2-club substructure. Formally, for a given non-overlapping 2-club substructure $M = (V_M, E_M)$ from $G = (V, E)$, the set of vertices to be added into $V_M$ (denoted as $AVS(V_M)$) is defined to be

$$AVS(V_M) = \{v_v | \forall v_x \in V_M, gd(v_x, v_v) \leq 2, and |N(v_v) \cap V_M|/|V_M| > 0.5, v_v \in N(M)\} \quad (2)$$

where $gd(v_x, v_v)$ representing the distance between vertices $v_x$ and $v_v$, the neighborhood sets of $M$ and $v_v$ are $N(M) = \{v_u | (v_u, v_v) \in E, v_v \in V_M, v_u \in V, v_u \notin V_M\}$, and $N(v_v) = \{v_u | (v_u, v_v) \in E, v_u \in V\}$.

Note that the new subnetwork on the vertex set $V_M \cap AVS(V_M)$ is still a 2-club substructure. This is because that any vertex in $AVS(V_M)$ must be of distance at most 2 to every vertex in $V_M$ and that every pair of vertices in $AVS(V_M)$ have a common neighbor in $V_M$. An example as shown in figure 3, the vertex $v_f$ is the overlapping vertex belonging to the two substructures $M_1$ and $M_2$ simultaneously since we have $v_f \in AVS(V_{M_1})$ and $v_f \in AVS(V_{M_2})$, where the vertices $v_g \notin AVS(M_1)$ since $|N(v_g) \cap V_{M_1}|/|V_{M_1}| < 0.5$, $v_h \notin AVS(M_1)$ since $gd(v_e, v_h) > 2$ and $v_m \notin AVS(M_1)$ since even $v_m \notin N(M_1) = \{v_f, v_g, v_h\}$.

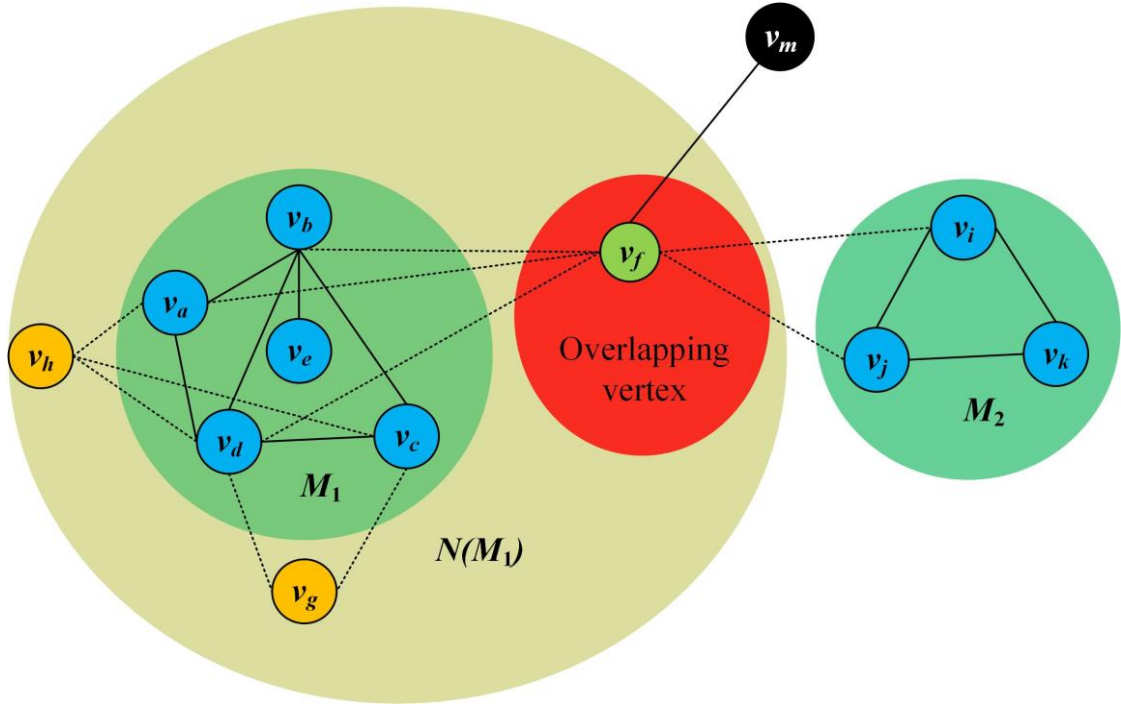

**Figure 3.** An example for demonstrating the 2-hop overlapping strategy. Vertex $v_f$ is the overlapping vertex searched by the 2-hop overlapping strategy, which belonging to the substructures $M_1$ and $M_2$.

*2.4.3. Details of DIVANC.* As shown in the following table 2, DIVANC removes edges iteratively according to their edge niche centrality until all the connected components are 2-club substructures. If needed, DIVANC can include an additional step to construct overlapping 2-club substructures by using the 2-hop overlapping strategy.

The effectiveness in practice of DIVANC will be reported in Section 3. In the following, we discuss its worst-case complexity. Let $\bar{k}$ be the average degree of the vertices and $T$ the number of edges removed. If the overlapping step is not performed, the time complexity of DIVANC is $O(\bar{k}^2|E|+\bar{k}^4 T)$ where the first term is the time to compute the edge niche centrality for all edges and the second term is the time to remove the $T$ edges. When the overlapping step is performed, the total running time is $O(\bar{k}^2|E|+\bar{k}^4 T+\bar{k}|V|)$. Since general practical networks usually have a small average degree $\bar{k}$ and since $T$ is at most the sum of $|E|-(|V|-1)$ and the number of detected 2-club substructures, the running time of DIVANC is very low, thus it is very efficiently. We note that neither $T$ nor the number of detected 2-club substructures is a parameter of the algorithm.

**Table 2.** The diagram of DIVANC.

| Algorithm DIVANC |
| --- |

**Input:** network $G=(V,E)$;

**Output:** 2-club substructures;

1: Calculate the edge niche centrality score for each edge of $G=(V,E)$;

2: *While* there are connected component of diameter greater than 2 *do*
3:     Remove the edge with the highest niche centrality score among all the edges;
4:     Re-calculate the scores of those edges affected by the removal of the edge;
5: *End while*
6: *If* overlapping 2-club substructures needed *then*
7:     *For* each of the current connected components *do*
8:         Apply the 2-hop overlapping strategy;
9:     *End for*
10: *End if*

## 3. Experiments and analyses

We report our experiment results and their analyses in this section. In Section 3.1 we compare DIVANC with existing algorithms in the literature on the PPI networks, LFR synthetic networks and football networks to verify the effectiveness of triad-rich substructures. We show the advantage of DIVANC in detecting sparse metadata groups in Section 3.2. In Section 3.3 we test the practical performance of our proposed edge niche centrality and 2-hop overlapping strategy.

*3.1. Verifying the effectiveness of triad-rich substructures*

In this section we mainly compare DIVANC with other widely-used reference algorithms to test the effectiveness of triad-rich substructures. To compare fairly, we select corresponding competing non-overlapping and overlapping algorithms respectively since DIVANC can also be extended into overlapping version, which is denoted as DIVANC' temporarily for comparison. Other than the well-known community detection algorithms, we also choose some excellent domain-specific algorithms (detecting protein complexes) such as COACH [43] and ClusterOne [5].

Among the non-overlapping algorithms, we freely downloaded the Cytoscape plugin for MCODE [4] at http://www.cytoscape.org/. We implemented INFOMAP [3] freely by the R package igraph [44]. We obtained the source code for MCL at http://www.micans.org/mcl. We obtained the code of MATLAB version of LOUVAIN [7] at http://perso.uclouvain.be/vincent.blondel/research/louvain.html. EPCA [11] for detecting the defined cograph communities based on the edge $P_4$ centrality, which is one critical component of the edge niche centrality proposed in this paper and we have the code. Especially, in order to verify the delicate advantages of edge niche centrality over than edge $P_4$ centrality, we scrabble up an especial edge division algorithm based on edge $P_4$ centrality to detect 2-club substructures (temporarily denoted as EPD2, Edge $P_4$ centrality and Diameter 2 stop criterion)) like our DIVANC. Thus, the different effectiveness of DIVANC and EPD2 can be just due to their own different edge centralities.

While, as for the overlapping algorithms, we freely downloaded the Cytoscape plugin for ClusterOne [5] at http://www.cytoscape.org/. We obtained the executable program for COACH [43] at http://www1.i2r.a-star.edu.sg/~xlli/. We used LinkComm [6] by its R package [45]. We made use of its fast version OSLOM2 [8] at http://oslom.org/software.htm. We set all the corresponding parameters of those competing algorithms at their respective default values as they report that the algorithms can obtain best performances under default parameter values. Especially inspired by scrabbling up the special algorithm EPD2 among the non-overlapping algorithms, in this section we further extend EPD2 into its overlapping version based the proposed 2-hop overlapping strategy, which is denoted as EPD2'. Introducing EPD2' as a competing algorithm can not only provide further comparative perspective between edge niche centrality and $P_4$ centrality in an overlapping context, but also can verify the portability of the 2-hop overlapping strategy.

The effectiveness of those algorithms are evaluated using a series of indices in terms of protein complex detection and GO term detection. We use the indices of the numbers of matching metadata groups, the cluster-wise sensitivity (*Sn*), cluster-wise positive predictive value (*PPV*), the accuracy score (*Acc*), maximum matching ratio (MMR) [5,43,46] to assess the algorithms in complexes detection. F-measure and Percentage of matched GO terms and MMR are used to assess them in identifying GO terms [34,35]. More details about the used indices can be found in Appendix B.

*3.1.1. Comparison in detecting protein complexes and GO terms on PPI networks.* The results on the effectiveness for detecting protein complexes of non-overlapping algorithms are summarized in table 3 and overlapping ones in table 4. Among the indices, we mainly pay more attention to the three indices: numbers of candidate complexes which can match at least one reference complex among golden standards (NMC), the accuracy scores (*Acc*) and maximum matching ratio (MMR), as given in bold fonts in tables 3 and 4. In addition to comparing them in detecting protein complexes, we also compare their effectiveness in detecting GO terms. We test the compared algorithms for detecting GO terms from *Sce*DIP and *Hsa*HPRD using the indices of F-measure, percentage of matched GO terms and maximum matching ratio. Figures 4(a-c) show the indices of F-measure, percentage of matched GO terms and maximum matching ratio for non-overlapping algorithms on *Sce*DIP and *Hsa*HPRD respectively. Figures 4(d-f) display the corresponding indices for overlapping algorithms.

As shown in table 3, among the non-overlapping algorithms, DIVANC has the largest numbers of matched protein complexes across all the golden standards except PCDq. Where it has 377 matched protein complexes, which is almost equal to the highest number 378. The maximum matching ratios of DIVANC are the highest one on SGD and CORUM, and while across MIPS and PCDq the maximum matching ratios of DIVANC are very close to the highest ones. The accuracy scores of DIVANC are also very close to their highest ones such as those of MCL, EPCA and EPD2. As demonstrated in table 4, among the overlapping algorithms, DIVANC' has the largest accuracy scores across all the golden standards. Except on PCDq DIVANC has the largest maximum matching ratio, on other golden standards the maximum matching ratios of DIVANC are lower than those of COACH. As figure 4 shows, the bar plots for illustrating the effectiveness in GO terms detection also clearly reveal that DIVANC, MCL are competitive among non-overlapping algorithms, while among overlapping algorithms DIVANC', COACH, LinkComm are competitive and they all outperform others like the instances about complexes detection described in tables 3 and 4. The reason for COACH and LinkComm possessing better effectiveness than DIVANC' in detecting GO terms is that both of COACH and LinkComm can obtain highly overlapping candidate metadata groups, while DIVANC' can just obtain periphery overlapping 2-club substructures. Thus in the further research on the one hand we should continue maintaining the unique graph-theoretic characteristics of 2-club substructures, on the other hand we should pay more attention to improving their overlapping extent.

To give a specific example, we especially select a simple complex named CCBL2-HBXIP-RABIF-UTP14A complex, which can be detected perfectly by DIVANC' but cannot by other algorithms. As shown in figure 5, the CCBL2-HBXIP-RABIF-UTP14A complex consists of four-subunit proteins, which is a protein complex stored in an integrated database of human genes and transcripts, the H-Invitational Database (H-InvD) [47]. The proteins in green color are the members of CCBL2-HBXIP-RABIF-UTP14A complex and those in dark red color are not. We emphasize that among all the non-overlapping and overlapping algorithms, only DIVANC' can detect the CCBL2-HBXIP-RABIF-UTP14A complex perfectly, while none of the algorithms MCODE, INFOMAP, LinkComm and OSLOM2 can detect meaningful candidate complex successfully, not to mention matching perfectly with the benchmark, more details in table C1.

**Table 3.** Comparison with non-overlapping algorithms for detecting complexes from *Sce*DIP and *Hsa*HPRD.

| Net[a] | GS[b] | Alg[c] | Cov[d] | NM[e] | AS[f] | NMC[g] | $Sn$ | $PPV$ | $Acc$ | $MMR$ |
|---|---|---|---|---|---|---|---|---|---|---|
| *Sce*DIP | MIPS | | 1061 | 203 | 12.52 | | | | | |
| | | MCODE | 781 | 51 | 15.31 | 15 | 0.2149 | 0.1987 | 0.2066 | 0.0301 |
| | | INFOMAP | 4980 | 441 | 11.29 | 47 | 0.4915 | 0.3190 | **0.3960** | 0.0961 |
| | | MCL | 4736 | 928 | 5.10 | 69 | 0.3125 | 0.3689 | 0.3395 | 0.1666 |
| | | LOUVAIN | 4980 | 675 | 7.38 | 35 | 0.5081 | 0.2571 | 0.3614 | 0.0849 |
| | | EPCA | 4687 | 1019 | 4.60 | 82 | 0.3530 | 0.3982 | 0.3749 | **0.2006** |
| | | EPD2 | 4723 | 1015 | 4.65 | 82 | 0.3589 | 0.3984 | 0.3782 | 0.1980 |
| | | DIVANC | 4856 | 1128 | 4.30 | **88** | 0.3526 | 0.4008 | 0.3759 | 0.2004 |
| | SGD | | 1211 | 305 | 5.70 | | | | | |
| | | MCODE | 781 | 51 | 15.31 | 21 | 0.3076 | 0.2490 | 0.2768 | 0.0358 |
| | | INFOMAP | 4980 | 441 | 11.29 | 74 | 0.6354 | 0.4447 | 0.5316 | 0.1050 |
| | | MCL | 4736 | 928 | 5.10 | 124 | 0.5026 | 0.5585 | 0.5298 | 0.1884 |
| | | LOUVAIN | 4980 | 675 | 7.38 | 79 | 0.6538 | 0.3598 | 0.4850 | 0.1259 |
| | | EPCA | 4687 | 1019 | 4.60 | 129 | 0.5348 | 0.5943 | **0.5638** | 0.2073 |
| | | EPD2 | 4723 | 1015 | 4.65 | 128 | 0.5382 | 0.5924 | 0.5647 | 0.2023 |
| | | DIVANC | 4856 | 1128 | 4.30 | **141** | 0.5348 | 0.5918 | 0.5626 | **0.2108** |
| *Hsa*HPRD | PCDq | | 3433 | 1204 | 4.51 | | | | | |
| | | MCODE | 1121 | 100 | 11.21 | 27 | 0.1624 | 0.1816 | 0.1717 | 0.0990 |
| | | INFOMAP | 9269 | 668 | 13.88 | 150 | 0.5192 | 0.3266 | 0.4118 | 0.0480 |
| | | MCL | 8903 | 1789 | 4.98 | 316 | 0.3992 | 0.5322 | 0.4609 | 0.1246 |
| | | LOUVAIN | 9269 | 1097 | 8.45 | 226 | 0.5385 | 0.2944 | 0.3981 | 0.0962 |
| | | EPCA | 8807 | 1946 | 4.53 | 377 | 0.3856 | 0.5504 | 0.4607 | **0.1450** |
| | | EPD2 | 8855 | 1942 | 4.56 | **378** | 0.3872 | 0.5491 | **0.4611** | 0.1448 |
| | | DIVANC | 9077 | 2151 | 4.22 | 377 | 0.3804 | 0.5587 | 0.4610 | 0.1443 |
| | CORUM | | 1955 | 1294 | 5.06 | | | | | |

| Alg | Cov | NM | AS | NMC | Sn | PPV | Acc | MMR |
|---|---|---|---|---|---|---|---|---|
| MCODE | 1121 | 100 | 11.21 | 23 | 0.2452 | 0.0791 | 0.1392 | 0.0087 |
| INFOMAP | 9269 | 668 | 13.88 | 73 | 0.5251 | 0.1591 | 0.2890 | 0.0210 |
| MCL | 8903 | 1789 | 4.98 | 190 | 0.4041 | 0.2460 | **0.3153** | 0.0613 |
| LOUVAIN | 9269 | 1097 | 8.45 | 95 | 0.5663 | 0.1310 | 0.2724 | 0.0327 |
| EPCA | 8807 | 1946 | 4.53 | 196 | 0.3772 | 0.2529 | 0.3088 | 0.0642 |
| EPD2 | 8855 | 1942 | 4.56 | 196 | 0.3810 | 0.2528 | 0.3103 | 0.0639 |
| DIVANC | 9077 | 2151 | 4.22 | **231** | 0.3735 | 0.2599 | 0.3116 | **0.0723** |

[a] Net Networks.

[b] GS Golden standards.

[c] Alg Algorithms.

[d] Cov Numbers of coverage proteins.

[e] NM Numbers of detected candidate complexes.

[f] AS Average size of obtained candidate complexes.

[g] NMC Numbers of candidate complexes which can match at least one reference complex.

**Table 4.** Comparison with overlapping algorithms for detecting complexes from *Sce*DIP and *Hsa*HPRD.

| Net[a] | GS[b] | Alg[c] | Cov[d] | NM[e] | AS[f] | NMC[g] | Sn | PPV | Acc | MMR |
|---|---|---|---|---|---|---|---|---|---|---|
| *Sce*DIP | MIPS | | 1061 | 203 | 12.52 | | | | | |
| | | COACH | 7814 | 886 | 8.82 | **165** | 0.3790 | 0.2781 | 0.3246 | **0.2831** |
| | | ClusterOne | 2218 | 596 | 3.72 | 76 | 0.2645 | 0.3815 | 0.3176 | 0.1424 |
| | | LinkComm | 6587 | 875 | 7.53 | 156 | 0.4176 | 0.3299 | 0.3711 | 0.2299 |
| | | OSLOM2 | 5442 | 85 | 64.02 | 21 | 0.5053 | 0.2382 | 0.3469 | 0.0310 |
| | | EPD2' | 4986 | 1015 | 4.91 | 89 | 0.3754 | 0.3845 | 0.3800 | 0.2121 |
| | | DIVANC' | 5129 | 1128 | 4.55 | 96 | 0.3896 | 0.3862 | **0.3879** | 0.2221 |
| | SGD | | 1211 | 305 | 5.70 | | | | | |
| | | COACH | 7814 | 886 | 8.82 | **232** | 0.5509 | 0.3774 | 0.4560 | **0.2731** |
| | | ClusterOne | 2218 | 596 | 3.72 | 126 | 0.4158 | 0.5941 | 0.4970 | 0.1767 |
| | | LinkComm | 6587 | 875 | 7.53 | 213 | 0.5543 | 0.3989 | 0.4703 | 0.2123 |
| | | OSLOM2 | 5442 | 85 | 64.02 | 24 | 0.6475 | 0.2745 | 0.4216 | 0.0298 |
| | | EPD2' | 4986 | 1015 | 4.91 | 133 | 0.5566 | 0.5629 | 0.5597 | 0.2107 |
| | | DIVANC' | 5129 | 1128 | 4.55 | 150 | 0.5716 | 0.5626 | **0.5671** | 0.2272 |
| *Hsa*HPRD | PCDq | | 3433 | 1204 | 4.51 | | | | | |
| | | COACH | 14086 | 1725 | 8.17 | **422** | 0.3937 | 0.1584 | 0.2497 | 0.1096 |
| | | ClusterOne | 4151 | 1103 | 3.76 | 286 | 0.2591 | 0.6746 | 0.4181 | 0.1029 |
| | | LinkComm | 11194 | 1605 | 6.97 | 330 | 0.3750 | 0.2958 | 0.3331 | 0.0826 |
| | | OSLOM2 | 10016 | 208 | 48.15 | 19 | 0.5262 | 0.1686 | 0.2978 | 0.0068 |
| | | EPD2' | 9300 | 1942 | 4.79 | 402 | 0.4088 | 0.4506 | 0.4292 | **0.1507** |
| | | DIVANC' | 9626 | 2151 | 4.48 | 401 | 0.4058 | 0.4564 | **0.4304** | **0.1507** |
| | CORUM | | 1955 | 1294 | 5.06 | | | | | |
| | | COACH | 14086 | 1725 | 8.17 | **443** | 0.4653 | 0.0681 | 0.1780 | **0.1103** |
| | | ClusterOne | 4151 | 1103 | 3.76 | 164 | 0.2711 | 0.2780 | 0.2745 | 0.0548 |
| | | LinkComm | 11194 | 1605 | 6.97 | 372 | 0.4308 | 0.1342 | 0.2404 | 0.0910 |
| | | OSLOM2 | 10016 | 208 | 48.15 | 20 | 0.5425 | 0.0970 | 0.2294 | 0.0057 |
| | | EPD2' | 9300 | 1942 | 4.79 | 213 | 0.4201 | 0.2113 | 0.2980 | 0.0697 |

|   |   |   |   |   |   |   |   |   |
|---|---|---|---|---|---|---|---|---|
| DIVANC' | 9626 | 2151 | 4.48 | 254 | 0.4233 | 0.2131 | **0.3004** | 0.0800 |

The foot note of table 4 being the same as that of table 3.

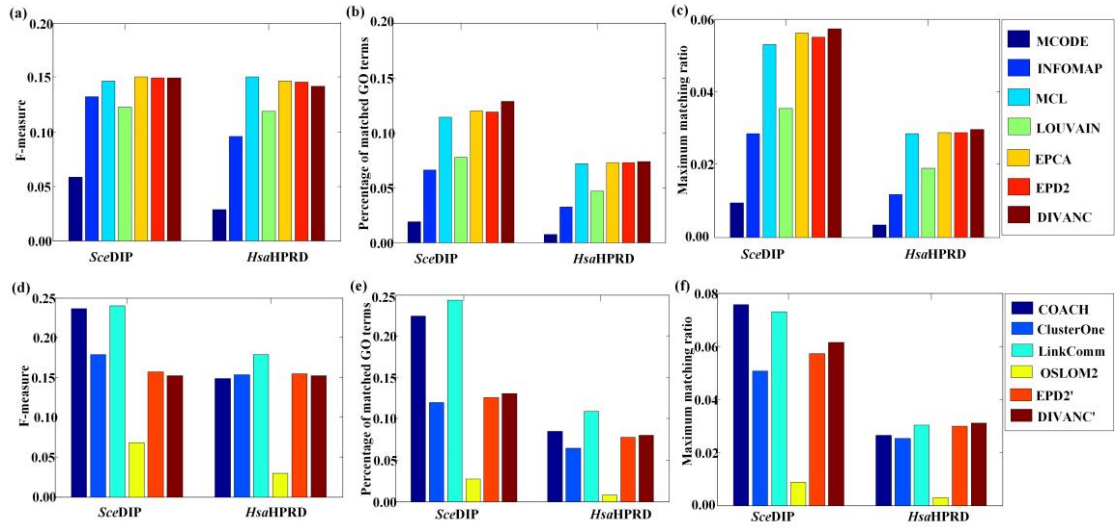

**Figure 4.** The bar plots illustrating the effectiveness of non-overlapping and overlapping algorithms for detecting GO terms. Figures 4(a-c) demonstrating the indices of F-measure, percentage of matched GO terms and maximum matching ratio of non-overlapping algorithms on *Sce*DIP and *Hsa*HPRD respectively; Figures 4(d-f) displaying the corresponding indices of overlapping algorithms.

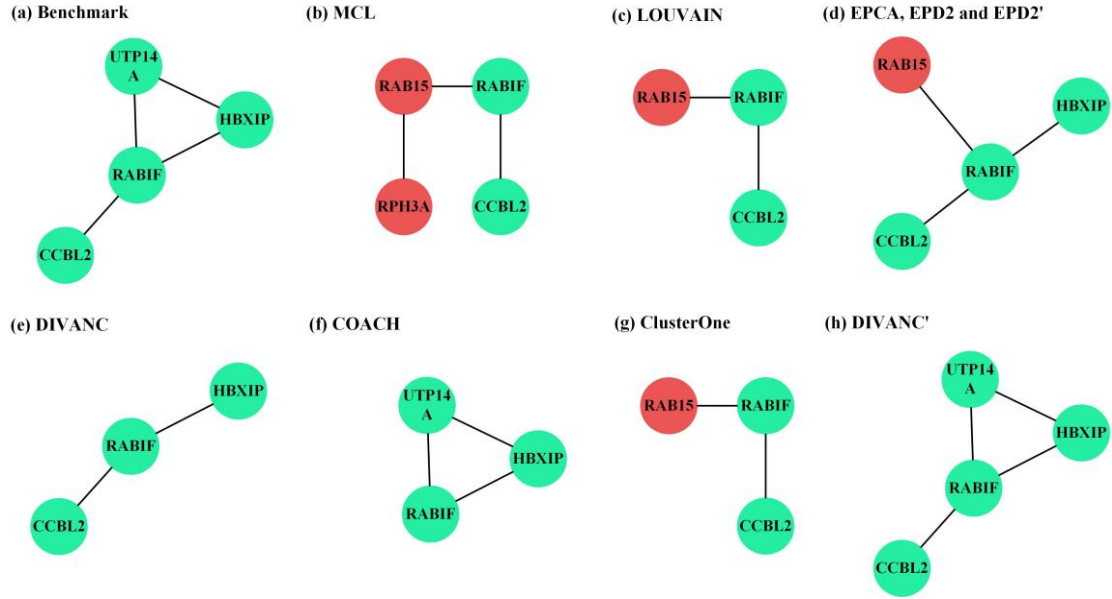

**Figure 5.** Illustration of the candidate complexes detected by the competing algorithms about CCBL2-HBXIP-RABIF-UTP14A complex. Figure 5(a) the benchmark of CCBL2-HBXIP-RABIF-UTP14A complex and figures 5(b-h) the corresponding candidate complexes detected by the non-overlapping algorithms MCL, LOUVAIN, EPCA, EPD2, DIVANC, and the overlapping algorithms COACH, ClusterOne, EPD2', DIVANC', where the genes in green color being the members of CCBL2-HBXIP-RABIF-UTP14A complex and those in dark red color not. Unfortunately, none of the algorithms MCODE, INFOMAP, LinkComm and OSLOM2

being able to detect valuable candidate complex successfully.

*3.1.2. Comparing the algorithms on LFR synthetic networks.* Although as the foundation of our 2-club substructures framework, the triad-rich substructures assumption about metadata groups are observed from PPI networks, what we want to emphasize is that either DIVANC or DIVANC' can work well on general complex networks. In the following we mainly test the scalability of DIVANC on LFR synthetic networks [28,29]. In the testing experiments, we use the well-known normalized mutual information (NMI) [48,49] (more details see in Appendix B) for evaluating community detection algorithms.

The testing LFR synthetic networks include a series of non-overlapping networks and overlapping networks respectively. The parameters for producing LFR non-overlapping and overlapping synthetic networks are introduced in Section 2.1. We compare the NMI values of the results obtained by the compared non-overlapping and overlapping algorithms on the synthetic networks as shown in figure 6. Each node of the figure corresponds to the average NMI value over 20 LFR networks produced on the same parameters. The NMI values of all algorithms decrease as the mixing parameter $mu$ increases. The reason is that community structures of the LFR networks become fuzzier and fuzzier, and thus are more difficult to be detected correctly as $mu$ increases. As figure 6(a) shows, the purple line with diamond signs represents the NMI value of DIVANC and figure 6(b) shows that the purple line with cross signs represents that of DIVANC'. Moreover, the results of the rest other algorithms are indicated by the corresponding color lines with corresponding signs as shown in figure 6. INFOMAP can obtain the best effectiveness among these compared non-overlapping algorithms and OSLOM2 has the highest NMI value among those overlapping algorithms. As figure 6 shows, DIVANC has competitive effectiveness among the non-overlapping algorithms and the second highest NMI value among overlapping algorithms. DIVANC obviously outperforms EPCA [11] and MCODE [4], while DIVANC' has better effectiveness than LinkComm [6], ClusterOne [5], and COACH [43]. Both of DIVANC and DIVANC' can obtain competitive effectiveness on LFR synthetic networks reveal us that the proposed 2-club substructure is suitable for synthetic networks at certain extent, but we really need to improve it since the triad-rich assumption is observed just only from PPI networks.

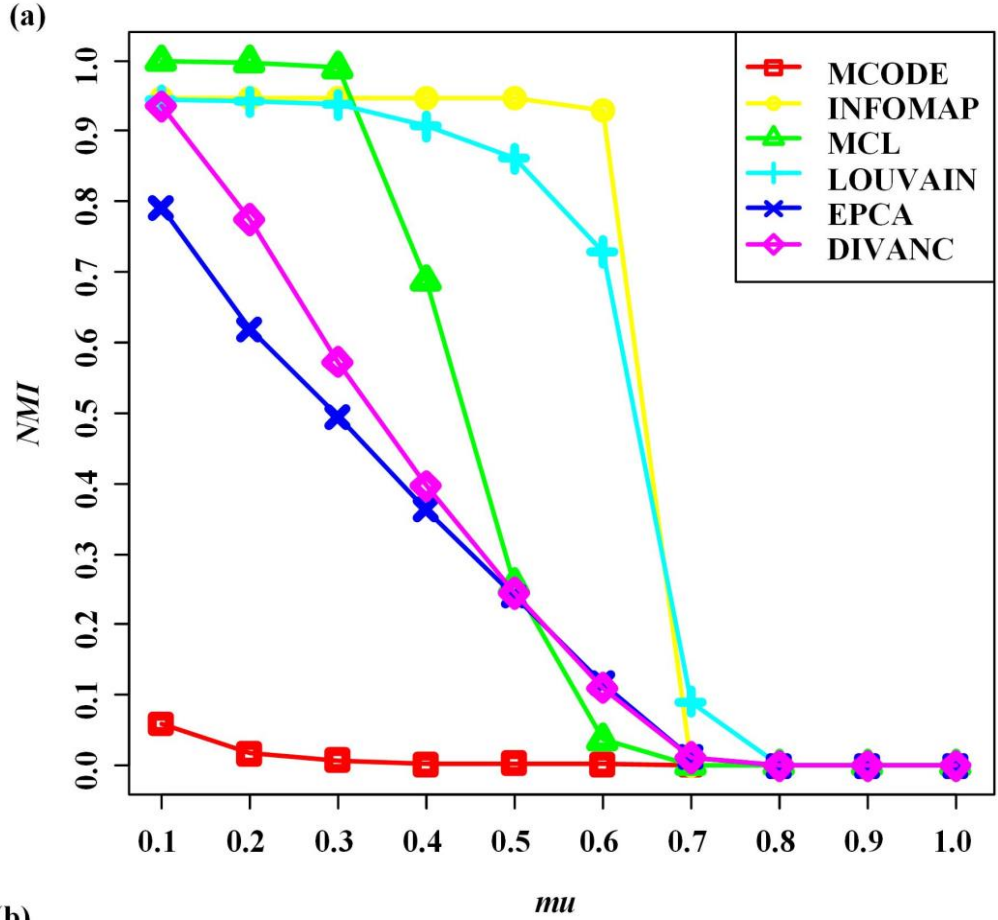
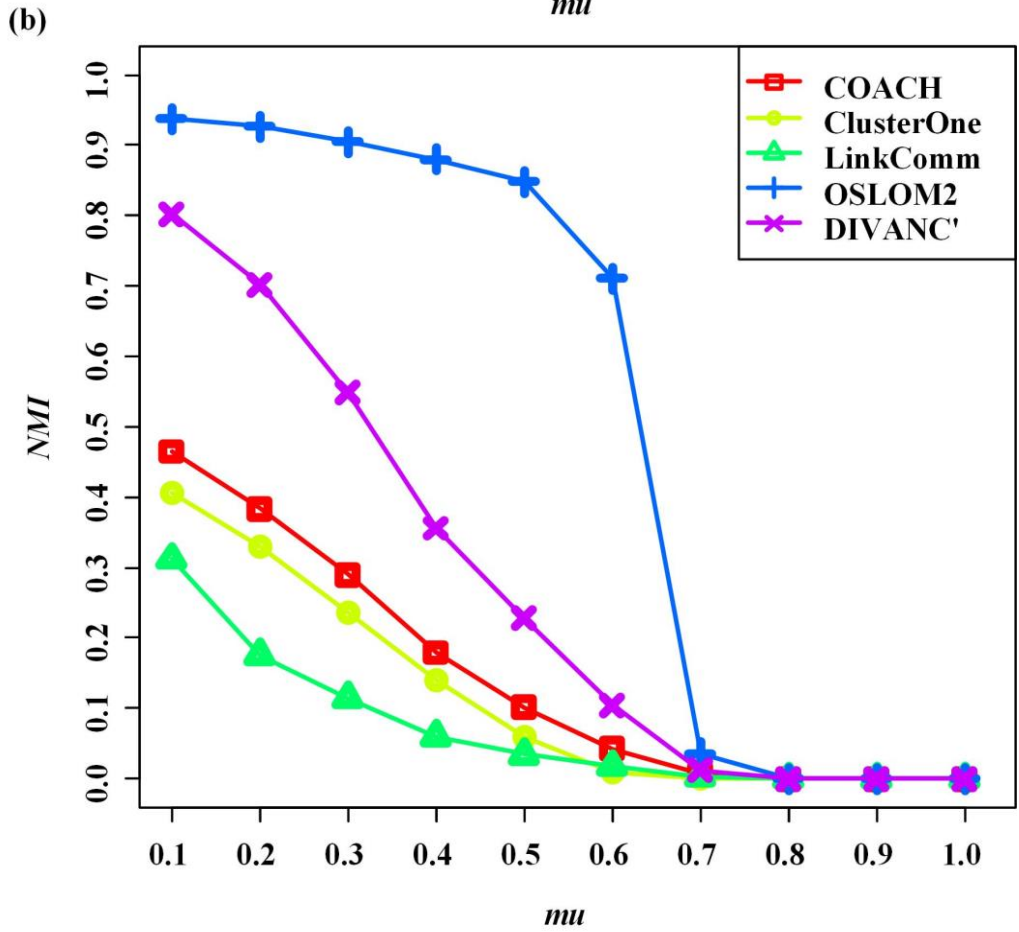

**Figure 6.** Illustration of the average NMI values of the results obtained by the compared algorithms on the series of LFR networks as *mu* from 0.1 to 1.0 with a step of 0.1. Figure 6(a) the average NMI values of non-overlapping algorithms; figure 6(b) the average NMI values of overlapping algorithms.

*3.1.3. Comparison on football networks.* In this section we also test them on a small social network the widely-used football networks [30,31]. As introduced in Section 2.1, football network consists of 115 teams and 613 games and the 115 teams are grouped into 11 conferences, with a 12th group of independent teams (without obvious affiliations, we artificially arrange the 8 independent teams into the 12th group together for convenience) as shown in figure 7(a). We display their NMI values of the compared non-overlapping and overlapping algorithms respectively in table 5. DIVANC and DIVANC' can obtain the same result. The NMI value of DIVANC is in close proximity to the highest one of LOUVAIN among non-overlapping algorithms, while among overlapping algorithms DIVANC' gains the highest NMI value. DIVANC gains 12 2-club substructures after removing 190 edges. Surprisingly, we find the 12 2-club substructures matching the 12 real football conferences in a nearly perfect way as shown in figure 7(b). Other than three of the 8 independent teams presented by green triangles as shown in figure 7(a) are misarranged just since they are the independent teams without obvious affiliations, all of the rest teams match the real groups perfectly. As shown in figure 7(b), two independent teams Navy and Notre Dame are arranged into the green circle group and another independent teams Connecticut is partitioned into the red triangle group irrelevantly. Notably DIVANC has signally better effectiveness than EPCA [11] since there are no isolated vertices among the obtained 2-club substructures, deleting 190 edges much lower than 290 ones that of EPCA and just only three misarranged teams, much fewer than that of EPCA. It is obvious that our algorithm also has impressive effectiveness on football networks.

**Table 5.** The NMI for effectiveness comparison on football networks.

| Non-overlapping algorithms | NMI of non-overlapping algorithms | Overlapping algorithms | NMI of overlapping algorithms |
|---|---|---|---|
| MCODE | 0.3834 | COACH | 0.5861 |
| INFOMAP | 0.8332 | ClusterOne | 0.6064 |
| MCL | 0.8332 | LinkComm | 0.2814 |
| LOUVAIN | **0.8361** | OSLOM2 | 0.8150 |
| EPCA | 0.6917 | DIVANC' | **0.8332** |
| DIVANC | 0.8332 | - | - |

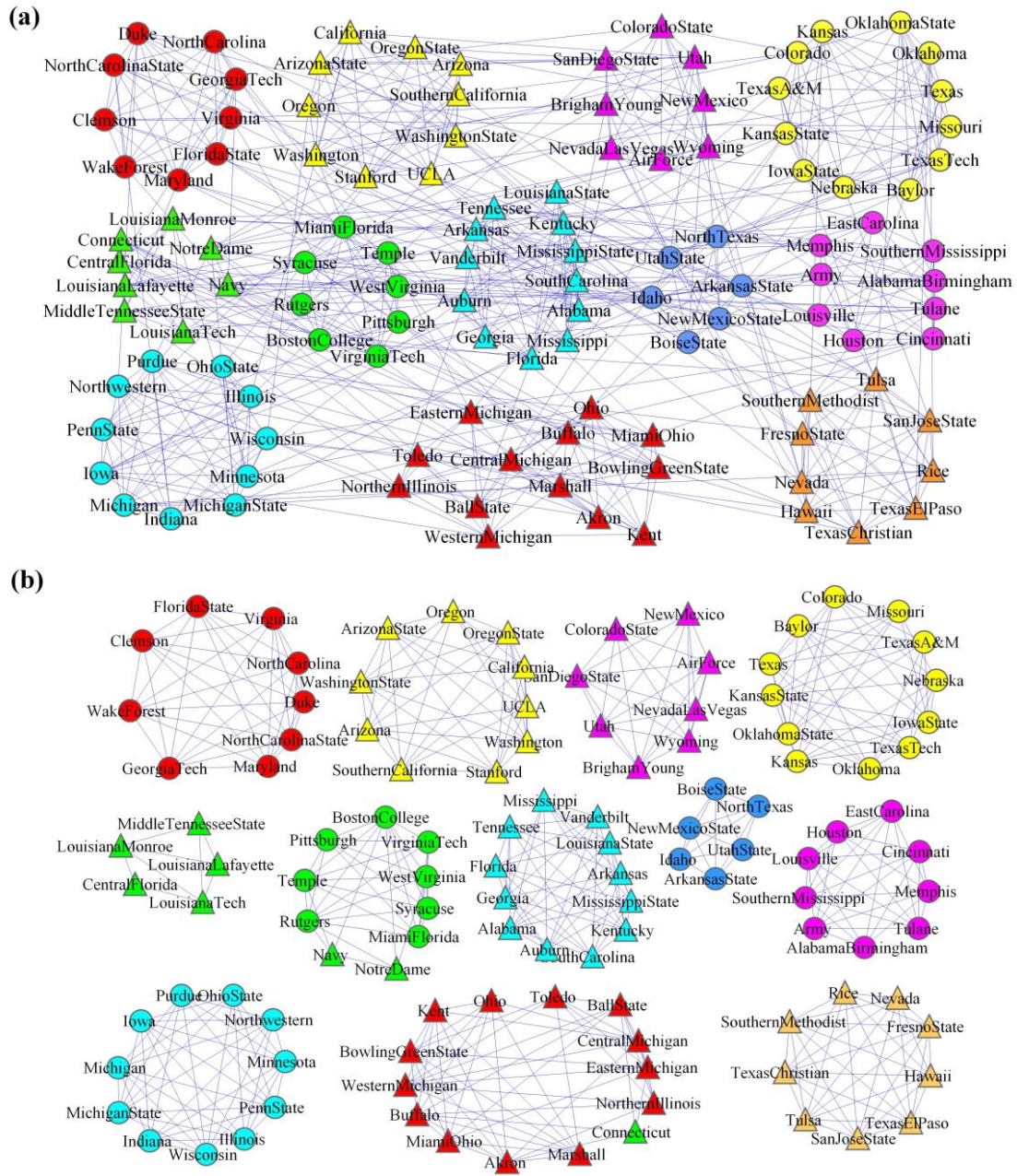

**Figure 7.** Illustration of the real groups and the 2-club substructures obtained by DIVANC on football networks. Figure 7(a) the football networks consisting of 12 groups; figure 7(b) the 2-club substructures obtained by DIVANC' after removing 190 edges.

*3.2. The advantage for detecting sparse metadata groups*

Other than the above macroscopic comparisons according various indices, in this section due to the triad-rich assumption that underpins our definition of 2-club substructures, we show the advantage of DIVANC for detecting sparse metadata groups. We list the details of 4 sparse benchmarks and their corresponding candidate metadata groups detected by the non-overlapping and overlapping algorithms in tables C2-C9 (Appendix C) and show them in figures 8-11. As described in tables C2 and C3, the density of GABAA receptor complex on *Hsa*HPRD is 0, thus it is really a challenge to detect its candidate complexes especially for the algorithms based on density. The algorithms based on density such as MCODE, COACH, ClusterOne and LinkComm even cannot obtain any valuable candidate ones which have common proteins with GABAA receptor complex. Among the algorithms, the neighborhood affinity scores (Appendix B) between the benchmark and the candidate complexes detected by DIVANC and DIVANC' are the highest. We also list the details about DGCR6L-ZNF193-ZNF232-ZNF446-ZNF446 complex in tables C4, C5 and display them in figure 9, eEF-1 complex in tables C6, C7 and in figure 10, the 116$^{th}$ complex of the golden standard MIPS in tables C8, C9 and in figure 11. The best effectiveness of DIVANC and DIVANC' in detecting sparse metadata groups again verifies the value of developing algorithms based on the triad-rich substructures.

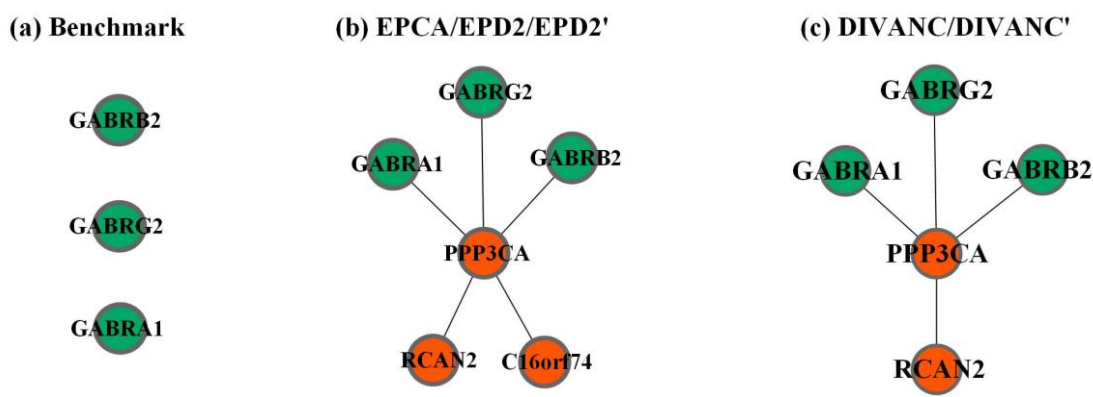

**Figure 8.** Illustration of the benchmark and candidate complexes detected by the competing algorithms about GABAA receptor complex. Figure 8(a) the benchmark of GABAA receptor complex; figures 8(b-c) the corresponding candidate complexes detected by the non-overlapping algorithms EPCA, EPD2, DIVANC and overlapping algorithms EPD2' and DIVANC'. The benchmark consisting of three isolated bright green proteins; among the detected candidate complexes the green proteins being the members of GABAA receptor complex and those in red color not.

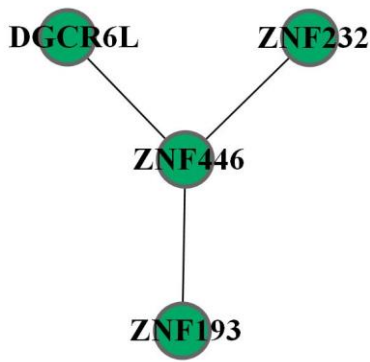
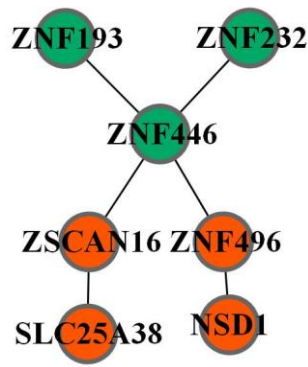
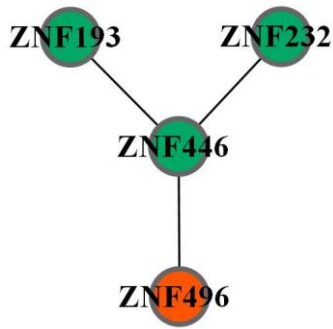
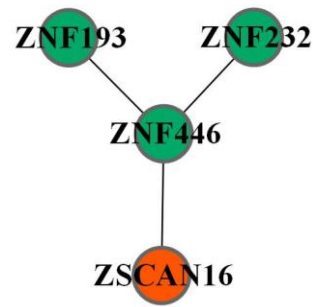
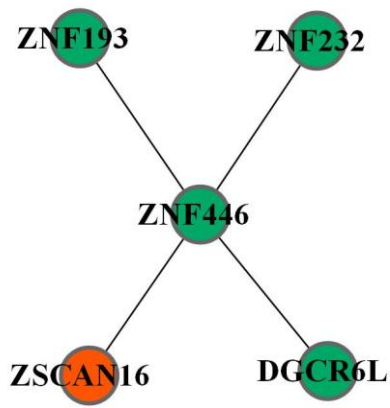

**Figure 9.** Illustration of the benchmark and candidate complexes detected by the competing algorithms about DGCR6L-ZNF193-ZNF232-ZNF446-ZNF446 complex. Figure 9(a) the benchmark of DGCR6L-ZNF193-ZNF232-ZNF446-ZNF446 complex; figures 9(b-e) the corresponding candidate complexes detected by the non-overlapping algorithms MCL, LOUVAIN, EPCA, EPD2, DIVANC and overlapping algorithms LinkComm, EPD2', DIVANC'. The benchmark consisting of 5 proteins coded by 4 genes, where the gene ZNF446 coded two proteins; among the detected candidate complexes the green color genes being the members of DGCR6L-ZNF193-ZNF232-ZNF446-ZNF446 complex and those in dark red color not.

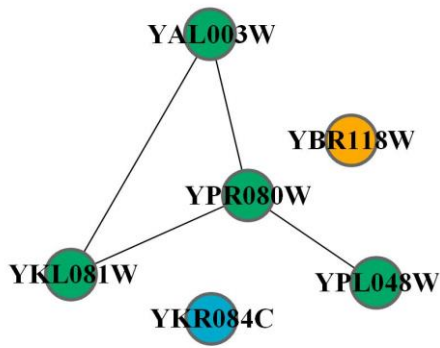
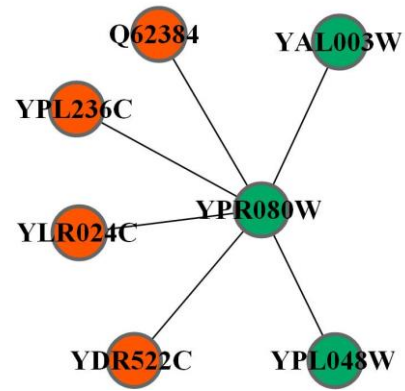

**Figure 10.** Illustration of the benchmark and candidate complexes by the competing algorithms about eEF-1 complex. Figure 10(a) the benchmark of eEF-1 complex; figure 10(b) the corresponding candidate complexes detected by DIVANC and DIVANC'. The benchmark consisting of 6 proteins, where the bright blue protein YKR084C is isolated proteins and YBR118W does not belong to the current input PPI networks since the incompleteness of datasets; among the detected candidate complexes the green proteins being the members of eEF-1 complex but those in dark red color not.

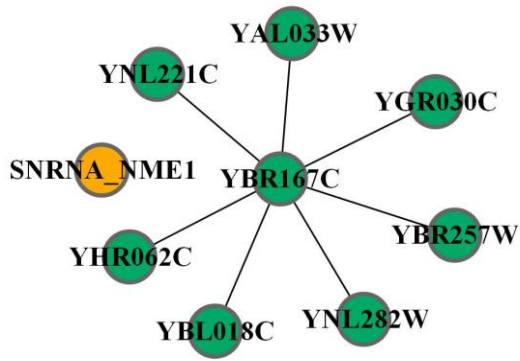
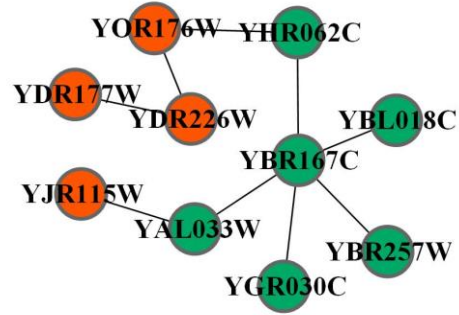
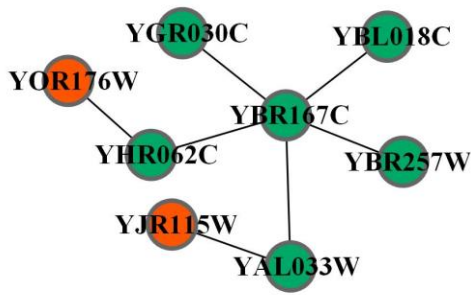
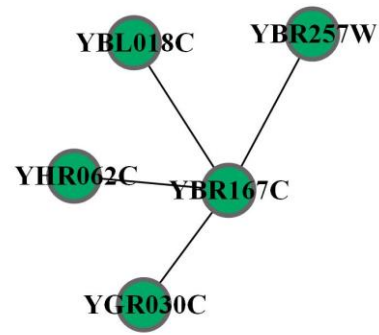
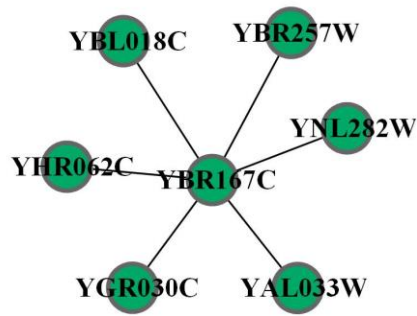

**Figure 11.** Illustration of the benchmark and candidate complexes detected by the competing algorithms about MIPS (116$^{th}$) complex. Figure 11(a) the benchmark of MIPS (116$^{th}$) complex; figures 11(b-e) the corresponding candidate complexes detected by INFOMAP, MCL, LOUVAIN, EPCA, EPD2, DIVANC and EPD2', DIVANC'. The benchmark consisting of 9 proteins, where the protein SNRNA_NME1 does not belong to the input PPI networks for the incompleteness of datasets; among the detected candidate complexes the green color proteins being the members of MIPS (116$^{th}$) complex but those in dark red color not.

*3.3. Testing practical performance of edge niche centrality and 2-hop overlapping strategy*
*3.3.1. Practical performance of edge niche centrality.* As we all know, edge centralities plays an important role in edge division algorithms, thus in this section we want to compare edge niche centrality with edge $P_4$ centrality to show its advantages since the former is developed based on the latter. In fact, the comparisons between edge niche centrality and edge $P_4$ centrality are able to be implemented into the comparisons among their corresponding algorithms. We compare DIVANC with EPD2 (as introduced in above, EPD2 is a special edge division algorithm consisting of edge $P_4$ centrality and diameter 2 stop criterion, just scrabbled up only in order to compare the practical performances of edge niche centrality with edge $P_4$ centrality in detecting 2-club substructures).

As displayed in tables 3 and 4, DIVANC detects 2151 candidate metadata groups while EPD2 obtains 1942 ones on *Hsa*HPRD, and DIVANC detects 1128 candidate metadata groups while EPD2 obtains 1015 ones on *Sce*DIP. Other than demonstrating their own relative indices of DIVANC and EPD2 in tables 3 and 4, we also display the differences between them in this section. We see that a detected candidate metadata group is able to match a golden standard complex or term if the score of neighborhood affinity (Appendix B) is equal or greater than 0.2 like in other parts of this paper. There are 249 candidate metadata groups detected by DIVANC which cannot match any one of the 1942 ones obtained by EPD2. In other word, there are 249 candidate metadata groups detected by DIVANC which cannot be obtained by EPD2 on *Hsa*HPRD, and likewise we can also detect 152 ones by DIVANC which cannot be obtained by EPD2 on *Sce*DIP. Surprisingly, among the 249 2-club substructures on *Hsa*HPRD, there are 71 ones which are able to match at least one metadata group, and 16 of the 152 ones on *Sce*DIP are able to match at least one metadata group. Further, some of the candidate ones detected by DIVANC that cannot be detected by EPD2 are even able to match at least one metadata group perfectly.

We display 8 candidate metadata groups detected by DIVANC but cannot by EPD2 from *Hsa*HPRD and their corresponding matched benchmarks in figure 12 and more details in table C10; 4 those candidate metadata groups from *Sce*DIP in figure 13 with more details in table C11. In addition to the official names of benchmarks and the gene members of benchmarks, we also list the scores of neighborhood affinity between the detected candidate metadata groups and their own benchmarks. In the columns of 'Candidate metadata groups' and 'Benchmark genes', we use bold fonts to label the common genes between candidate metadata groups and benchmark genes. In figures 12 and 13, each benchmark consists of the genes in the area circled by dotted line and the candidate metadata groups represented by the components consisting of the genes with red and green colors together. Among the genes in circled areas, the green genes are the common ones of candidate metadata groups and benchmarks. While, the bright blue, yellow and purple genes are the ones which cannot be detected by DIVANC. Notably, the yellow genes are the ones which do not belong to the used PPI networks temporarily for the incompleteness of datasets and the bright blue ones are isolated proteins from the PPI networks, thus they will never be are able to be detected by any algorithms. Only the purple ones are those missed by DIVANC. As we can see in figures 12 and 13, the genes of benchmarks cannot always be constructed as connected subnetworks also for the incompleteness property of the current PPI networks temporally. The fact that metadata groups within the networks which are not always connected subnetworks and not to mention dense subnetworks, is just the challenges for metadata groups detection. In a word, those practical effectiveness comparisons between DIVANC and EPD2 reveal an obvious advantage of edge niche centrality over edge $P_4$ centrality.

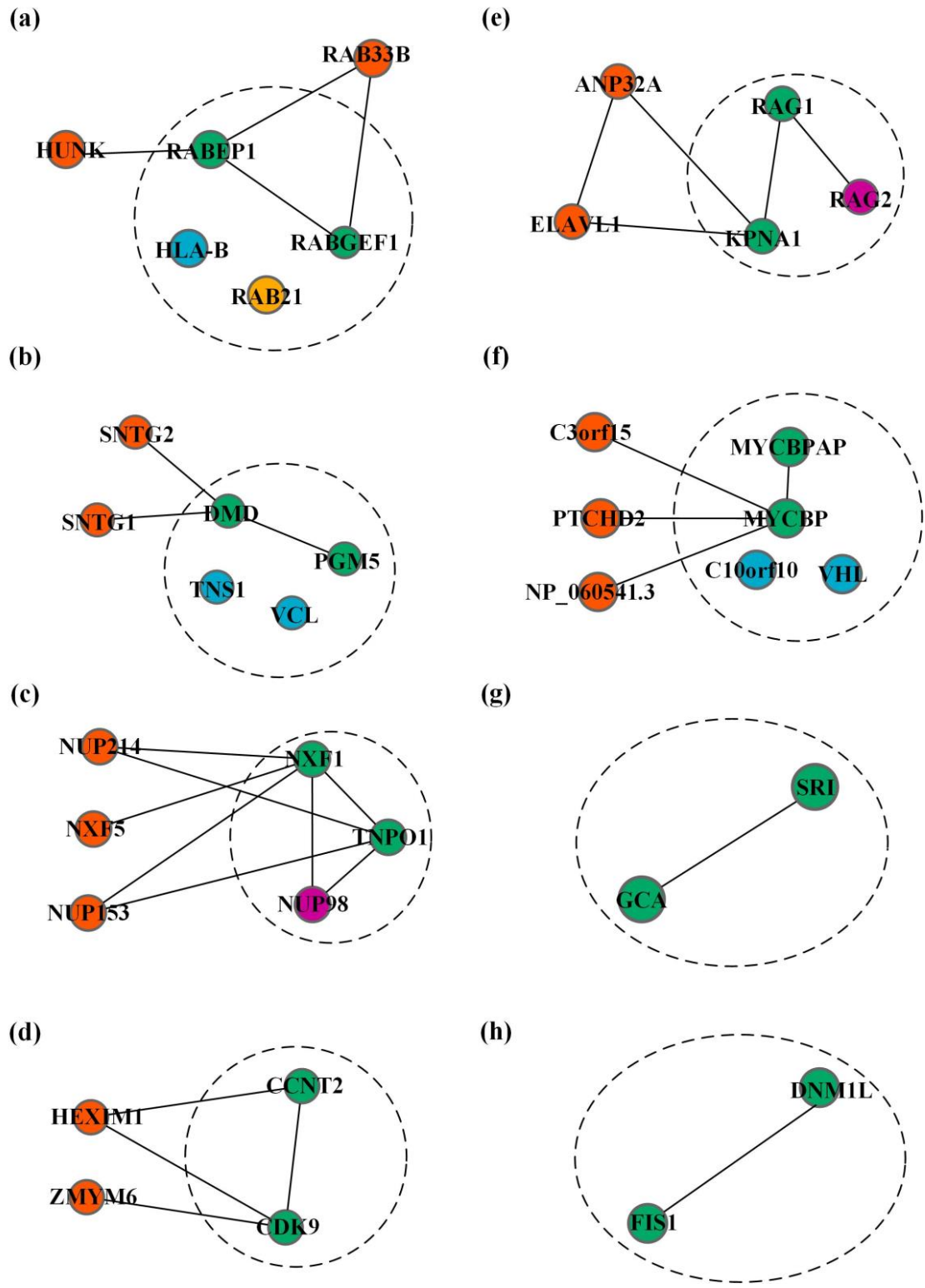

**Figure 12.** Illustration of the candidate metadata groups detected by DIVANC but cannot by EPD2 from *Hsa*HPRD. Figures 12(a-h) the 8 detected candidate metadata groups (the connected subnetworks consisting of green and red proteins) and the benchmarks in the areas circled by dotted line as listed in table C10.

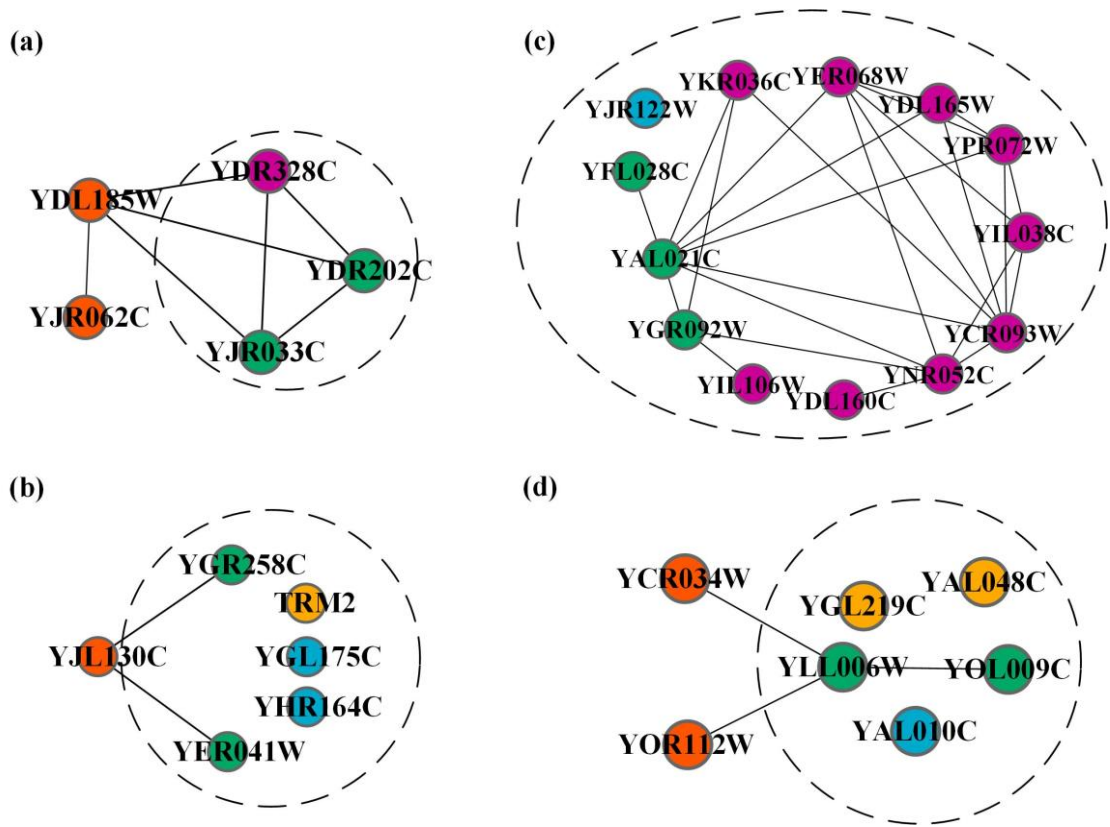

**Figure 13.** Illustration of the candidate metadata groups detected by DIVANC but cannot by EPD2 from *Sce*DIP. Figures 13(a-d) the 4 detected candidate metadata groups (the connected subnetworks consisting of green and red proteins) and the benchmarks in the areas circled by dotted line as listed in table C11.

*3.3.2. Performance of 2-hop overlapping strategy.* The algorithm DIVANC can be extended into overlapping version DIVANC' by the proposed 2-hop overlapping strategy. In this section we mainly test the performance of 2-hop overlapping strategy in detail since it plays the role in detecting overlapping 2-club substructures. As described in tables 3 and 4, whether on *Hsa*HPRD or on *Sce*DIP, DIVANC' performs better than DIVANC overall. In other words, the better effectiveness of DIVANC' justifies the value of the 2-hop overlapping strategy. The proposed 2-hop overlapping strategy not only produces new candidate metadata groups which can match metadata groups, but also can improve the matching levels between detected candidate metadata groups and their own benchmarks, and even makes some candidate ones to match benchmarks perfectly. Here we list 6 candidate metadata groups detected by DIVANC which are further improved by the 2-hop overlapping strategy to match their own benchmarks perfectly in figure 14. The genes with green color are those detected by DIVANC, while the genes with red color are those detected additionally by the 2-hop overlapping strategy. Thus the overlapping algorithm DIVANC' with 2-hop overlapping strategy can detect the candidate ones consisting of green genes and red genes together. As we list the neighborhood affinity scores in table C12, the candidate ones detected by DIVANC' can match their own benchmarks perfectly. Although we demonstrate the significant performance of the 2-hop overlapping strategy mainly by comparing the results of DIVANC and DIVANC', the improved results of EPD2' from EPD2 again verify the effects of 2-hop overlapping strategy as described in tables 3 and 4 from another point of view. Thus the promotional effectiveness of EPD2' over EPD2 also shows very well that the proposed 2-hop overlapping strategy has strong portability and can be widely used to turn other non-overlapping algorithms into overlapping ones.

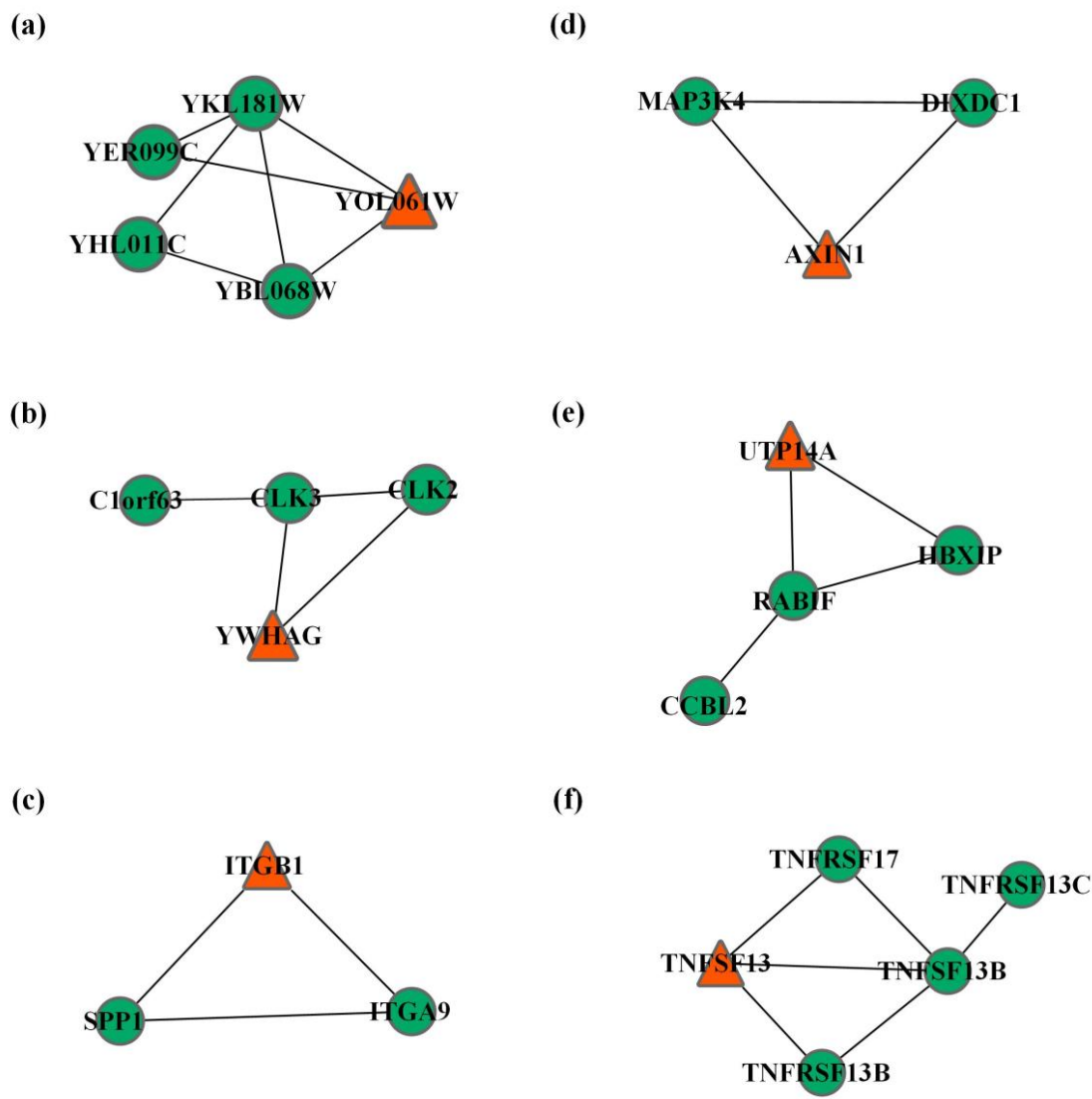

**Figure 14.** Illustration of the metadata groups detected by DIVANC' matching their own benchmarks perfectly. Figures 14(a-f) the 6 metadata groups as listed in table C12, where the red triangle proteins being those searched by the 2-hop overlapping strategy and together with the non-overlapping green circle proteins matching their own benchmarks perfectly.

### 4. Conclusions and discussion

In this work, we aim to overcome the challenge that traditional structural community definitions cannot characterize intrinsic features of metadata groups comprehensively. We develop a new framework by incorporating the novel assumption of triad-rich substructures, defining 2-club substructures, designing the effective algorithm DIVANC to detect non-overlapping and overlapping candidate metadata groups that have desired graph-theoretic properties. To verify the effectiveness of triad-rich substructures, we compare DIVANC with existing algorithms on PPI networks, LFR synthetic networks and football networks. The experimental results reveal DIVANC outperforms most other exiting algorithms significantly and, in particular, can detect sparse metadata groups.

In a future study, we will attempt to study the possible applications of 2-club subclasses on complex networks from the viewpoint of graph theory since 2-club substructures have interesting internal structures.

**Acknowledgments**

This work is supported by the National Natural Science Foundation of China (NSFC) under Grant numbers 61532014, 91530113, 61432010, 61303122, 61303118, 61402349, 71401130; and the Fundamental Research Funds for the Central Universities under Grant numbers BDZ021404.

**Appendix A. The details about two protein complexes**

*Nuclear origin of replication recognition complex*
A multisubunit complex that is located at the replication origins of a chromosome in the nucleus [20].

*GID complex*
A protein complex with ubiquitin ligase activity that is involved in proteasomal degradation of fructose-1,6-bisphosphatase (FBPase) and phosphoenolpyruvate carboxykinase during the transition from gluconeogenic to glycolytic growth conditions [21].

**Appendix B. The details about relative indices**

*The numbers of matching metadata groups*
The numbers of detected candidate metadata groups matching at least one metadata groups are important indices for comparison. A detected candidate metadata group $M_A$ with $|V_{M_A}|$ proteins or genes is thought to match with a metadata group $M_B$ with $|V_{M_B}|$ proteins or genes if the score of neighborhood affinity

$$NA(M_A, M_B) = |V_{M_A} \cap V_{M_B}|^2 \Big/ \left( |V_{M_A}| \times |V_{M_B}| \right) \geq \omega, \tag{B.1}$$

where the threshold $\omega$ is usually set as 0.2 or 0.25 in the references [43,46].

*Accuracy score*
We mainly make use of accuracy score ($Acc$) to evaluate the performance of various algorithms on protein complex detection. It is the geometric mean of cluster-wise sensitivity ($Sn$) and cluster-wise positive predictive value ($PPV$) [46]. Given $r$ detected and $s$ reference complexes, let $t_{ij}$ represent the number of proteins that existing in both detected complex $i$ and reference complex $j$, and $w_j$ represents the number of proteins in reference complex $j$. Then $Sn$ and $PPV$ are defined as $Sn = \sum_{j=1}^{s} \max_{i=1,\cdots,r} \{t_{ij}\} \Big/ \sum_{j=1}^{s} w_j$ ,

$$PPV = \sum_{i=1}^{r} \max_{j=1,\cdots,s} \{t_{ij}\} \bigg/ \sum_{i=1}^{r} \sum_{j=1}^{s} t_{ij}$$ respectively. Since $Sn$ can reach its maximum by grouping all proteins in one complex, whereas $PPV$ can be maximized by putting each protein in its own complex, we use their geometric mean

$$Acc = \sqrt{Sn \times PPV},\tag{B.2}$$

as 'accuracy' to balance these two indices [43,46], where the higher $Acc$ scores mean the better results.

*F-measure*

To investigate the performance of competing algorithms in detecting GO terms, we can compute the indices of F-measure [50]. If the neighborhood affinity score between a detected candidate GO term $p$ and a real GO term $rg$, $NA(p,rg) \geq \omega$, they are considered to be matched with each other. Assuming $PC$ as the set of candidate ones detected by an algorithm, $RG$ as the set of real GO terms, $N_{cp}$ indicating the number of candidate ones which can match at least one real GO term and $N_{crg}$ representing the number of real GO terms that match at least one candidate term, we obtain precision ($P$) and recall ($R$) as follows [50]: $N_{cp} = |\{p | p \in PC, \exists rg \in RG, NA(p,rg) \geq \omega\}|$,

$N_{crg} = |\{rg | rg \in RG, \exists p \in PC, NA(p,rg) \geq \omega\}|$, $P = N_{cp}/|PC|$, $R = N_{crg}/|RG|$. F-measure ($F$) as the harmonic mean of precision and recall, thus we have

$$F = (2 \times P \times R)/(P + R).\tag{B.3}$$

*Percentage of matched GO terms*

Percentage of matched GO terms which are considered to be the percentage of the GO terms which are correctly matched to at least one of the identified candidate GO terms [34,35].

*Maximum matching ratio*

Here we also use a measure called maximum matching ratio (MMR) [5] to evaluate relative algorithms on detection of protein complexes and GO terms. The MMR builds on maximal matching in a bipartite network, in which the two sets of vertices represent the reference and detected community, respectively, and an edge connecting a reference community with a detected one is weighted by the score of neighborhood affinity introduced in (equation B.1). We select the maximum weighted bipartite matching on this network; that is, we chose a subset of edges such that each detected and reference communities is incident on at most one selected edge and the sum of the weights of such edges is maximal. The chosen edges then represent an optimal assignment between reference and detected communities such that no reference community is assigned to more than one detected community and vice versa. The MMR between the detected and the reference community set is then given by the total weight of the selected edges, divided by the number of reference communities. MMR offers a natural, intuitive way to compare detected communities with a gold standard and it explicitly penalizes cases when a reference community is split into two or more parts in the predicted set, as only one of its parts is allowed to match the correct reference community.

*Normalized mutual information*

Normalized mutual information (NMI) is well known for evaluating community detection algorithms. In this paper we use the version of $NMI_{MGH}$ [48] to assess the similarities between detected results and golden standards on football networks and the series of LFR synthetic networks. Its definition is demonstrated as

$$NMI_{MGH} = I(X:Y)/\max(H(X), H(Y)), \qquad (B.4)$$

where $I(X:Y)$ is the mutual information, $H(X)$, ($H(Y)$) the unconditional entropy of cover $X$, ($Y$). More details can be found in the original references [48,49].

**Appendix C. Supplementary relative tables for more details**

In tables C1-C9, we mainly list the names and member genes of benchmark complexes, the common genes of candidate metadata groups with bold font; we also list the scores of neighborhood affinity, the sizes, density and whether the candidate metadata groups can match the benchmark complexes.

Table C1. The details of detection about CCBL2-HBXIP-RABIF-UTP14A complex.

| Benchmark/ Algorithms | Genes | Score | Size | Density | Success |
|---|---|---|---|---|---|
| CCBL2-HBXIP-RABIF-UTP14A complex | **RABIF HBXIP CCBL2 UTP14A** | - | 4 | 0.6667 | - |
| MCODE | - | - | - | - | - |
| INFOMAP | RAB3B RPH3AL SYTL4 RAB8A MAP4K2 RIMS2 RPH3A **RABIF** RAB3IP RAB3A RAB27A RAB3D RAB27B RAB3C RAB15 RABGGTB RAB7A MYO5A CHM RILP RNF115 TRIM2 PTP4A2 ALAS2 CHML SYTL5 RAB3IL1 DMXL2 **CCBL2** MLPH UNC13D SUCLA2 | 0.0313 | 32 | 0.1149 | NO |
| MCL | RPH3A **RABIF** RAB15 **CCBL2** | 0.2500 | 4 | 0.5000 | YES |
| LOUVAIN | **RABIF** RAB15 **CCBL2** | 0.3333 | 3 | 0.6667 | YES |
| EPCA | **RABIF** RAB15 **HBXIP CCBL2** | 0.5625 | 4 | 0.5000 | YES |
| EPD2 | **RABIF** RAB15 **HBXIP CCBL2** | 0.5625 | 4 | 0.5000 | YES |
| DIVANC | **RABIF HBXIP CCBL2** | 0.7500 | 3 | 0.6670 | YES |
| COACH | **UTP14A RABIF HBXIP** | 0.7500 | 3 | 1.0000 | YES |
| ClusterOne | **RABIF CCBL2** RAB15 | 0.3333 | 3 | 0.6667 | YES |
| LinkComm | SAT1 DLEU1 RNF10 **UTP14A** ZNF24 CCDC106 TRDMT1 PHF10 PDCD5 HNRNPH3 | 0.0250 | 10 | 0.2000 | NO |
| OSLOM2 | RAB25 RAB14 RAB3B RPH3AL SYTL4 GDI2 RAB8A RAB11A OPTN MAP4K2 RIMS2 RPH3A **RABIF** RAB3IP OCRL RAB11B RAB3A RAB27A RAB3D RAB27B RAB3C RAPGEF4 RAB15 RABGGTB MYO5B RAB7A RAB1A RAB6A BICD2 MYO5A TRIM3 CHM RAB11FIP2 RAB11FIP5 DYNLL2 TRIM2 RAB32 PTP4A2 ALAS2 CHML SYTL5 RAB3IL1 DMXL2 ADCK1 ZNF593 RAB1B **CCBL2** SSX2 RAB11FIP1 GOLGA1 GOLGA4 RABGAP1 KIF20A PMM1 KIAA1549 RAB11FIP3 SSR2 SRPR RUSC2 ARF5 RAB11FIP4 ARF4 MLPH UNC13D WDR44 MICALCL MICAL3 TROVE2 SUCLA2 MBD5 DNAL4 | 0.0141 | 71 | 0.0471 | NO |
| EPD2' | **RABIF** RAB15 **HBXIP CCBL2** | 0.5625 | 4 | 0.5000 | YES |
| DIVANC' | **RABIF HBXIP CCBL2 UTP14A** | 1.0000 | 4 | 0.6667 | YES |

**Table C2.** The details about GABAA receptor complex detected by non-overlapping algorithms.

| Benchmark/ Algorithms | Genes | Score | Size | Density | Success |
|---|---|---|---|---|---|
| **GABAA receptor complex** | **GABRA1 GABRB2 GABRG2** | - | 3 | 0.0000 | - |
| MCODE | - | - | - | - | - |
| INFOMAP | ZDHHC3 **GABRG2 GABRA1** PPP3CA C16orf74 HOXB2 CCS KARS SOD1 RCAN2 | 0.1333 | 10 | 0.2000 | NO |
| MCL | **GABRA1** PPP3CA C16orf74 RCAN2 | 0.0833 | 4 | 0.5000 | NO |
| LOUVAIN | DRD4 **GABRA1** | 0.1667 | 2 | 1.0000 | NO |
| EPCA | **GABRG2** PPP3CA **GABRA1** C16orf74 **GABRB2** RCAN2 | 0.5000 | 6 | 0.3333 | YES |
| EPD2 | **GABRG2** PPP3CA **GABRA1** C16orf74 **GABRB2** RCAN2 | 0.5000 | 6 | 0.3333 | YES |

| | DIVANC | **GABRG2** PPP3CA **GABRA1** **GABRB2** RCAN2 | 0.6000 | 5 | 0.4000 | YES |

**Table C3.** The details about GABAA receptor complex detected by overlapping algorithms.

| Benchmark/ Algorithms | Genes | Score | Size | Density | Success |
|---|---|---|---|---|---|
| **GABAA receptor complex** | **GABRA1 GABRB2 GABRG2** | - | 3 | 0.0000 | - |
| COACH | - | - | - | - | - |
| ClusterOne | - | - | - | - | - |
| LinkComm | - | - | - | - | - |
| OSLOM2 | TIAM1 SRC PRKCD PDPK1 PRKCA GRIN2B DAB2 SGSM3 ITGB2 MARCKS PRKCG PRKCB MIB1 DIP2A PRKD1 GNB2L1 CISH PRKCE PEBP1 C1QBP PRKCZ GJA1 PRKCI PARD6A HABP4 RASGRP3 GRM5 PARD3 NRGN PIK3CB PRKCH NFATC1 NUMB CASR PA2G4 CYTH2 CHAT ADAP1 AFAP1 PDLIM7 PPP1R14A **GABRG2** ADRBK1 ANXA7 CDCP1 PDE6G MT-ND2 FARP2 ANXA1 ITGB7 RIPK4 AKT3 **GABRA1** PDLIM5 RPS6KB2 MARK4 PARD6B ST7 GJA3 GCNT1 LST1 MAP2K5 HMGXB4 CSNK1G3 CSNK1G1 PNMA1 GABRA4 MSI1 PNMA6A LENG1 G2E3 C19orf29OS | 0.0185 | 72 | 0.0786 | NO |
| EPD2' | **GABRG2** PPP3CA **GABRA1** C16orf74 **GABRB2** RCAN2 | 0.5000 | 6 | 0.3333 | YES |
| DIVANC' | **GABRG2** PPP3CA **GABRA1 GABRB2** RCAN2 | 0.6000 | 5 | 0.4000 | YES |

**Table C4.** The details about DGCR6L-ZNF193-ZNF232-ZNF446-ZNF446 complex detected by non-overlapping algorithms.

| Benchmark/ Algorithms | Genes | Score | Size | Density | Success |
|---|---|---|---|---|---|
| **DGCR6L-ZNF193-ZNF232-ZNF446-ZNF446 complex** | ZNF446 ZNF446 DGCR6L ZNF232 ZNF193 | - | 5 | 0.3000 | - |
| MCODE | - | - | - | - | - |
| INFOMAP | **ZNF446** ZNF263 MZF1 LSM10 LSM11 ZNF473 SLBP SCAND1 ZNF434 ZNF202 ZNF167 ZSCAN20 ZSCAN21 TRIM41 **ZNF193 ZNF232** ZSCAN16 SLC25A38 | 0.1000 | 18 | 0.1373 | NO |
| MCL | NSD1 **ZNF446** ZNF496 **ZNF193 ZNF232** ZSCAN16 SLC25A38 | 0.2571 | 7 | 0.2857 | YES |
| LOUVAIN | **ZNF446** ZNF496 **ZNF193 ZNF232** | 0.4500 | 4 | 0.5000 | YES |
| EPCA | **ZNF446 ZNF193 ZNF232** ZSCAN16 | 0.4500 | 4 | 0.5000 | YES |
| EPD2 | **ZNF446 ZNF193 ZNF232** ZSCAN16 | 0.4500 | 4 | 0.5000 | YES |
| DIVANC | **ZNF446 DGCR6L ZNF193 ZNF232** ZSCAN16 | 0.6400 | 5 | 0.4000 | YES |

**Table C5.** The details about DGCR6L-ZNF193-ZNF232-ZNF446-ZNF446 complex detected by overlapping algorithms.

| Benchmark/ Algorithms | Genes | Score | Size | Density | Success |
|---|---|---|---|---|---|
| **DGCR6L-ZNF193-ZNF232-ZNF446-ZNF446 complex** | **ZNF446 ZNF446 DGCR6L ZNF232 ZNF193** | - | 5 | 0.3000 | - |
| COACH | SUFU **ZNF446** ZNF263 | 0.0667 | 3 | 1.0000 | NO |
| ClusterOne | KCNRG **DGCR6L** NP_612377.3 | 0.0667 | 3 | 0.6667 | NO |
| LinkComm | **ZNF446 ZNF193 ZNF232** ZSCAN16 | 0.4500 | 4 | 0.5000 | YES |
| OSLOM2 | RNF41 USP8 ZNF24 **ZNF446** ZNF263 ZNF496 CCDC130 MZF1 ZNF165 LSM10 FAM124A LSM11 ZNF473 SLBP ZNF174 SCAND1 ZNF434 ZNF202 ZNF167 ZSCAN20 ZSCAN21 TRIM41 **ZNF193 ZNF232** ZSCAN16 SLC25A38 | 0.0692 | 26 | 0.1015 | NO |
| EPD2' | **ZNF446 ZNF193 ZNF232** ZSCAN16 | 0.4500 | 4 | 0.5000 | YES |
| DIVANC' | **ZNF446 DGCR6L ZNF193 ZNF232** ZSCAN16 | 0.6400 | 5 | 0.4000 | YES |

**Table C6.** The details about eEF-1 complex detected by non-overlapping algorithms.

| Benchmark/ Algorithms | Genes | Score | Size | Density | Success |
|---|---|---|---|---|---|
| **eEF-1 complex** | **YPL048W YKL081W YPR080W YBR118W YKR084C YAL003W** | - | 6 | 0.2667 | - |
| MCODE | - | - | - | - | - |

| | INFOMAP | YLR456W YPR172W **YKR084C** YNL001W YNL105W YKR010C YBL033C | 0.0238 | 7 | 0.2857 | NO |
| --- | --- | --- | --- | --- | --- | --- |
| | MCL | **YPR080W** YEL034W YKL085W **YPL048W** Q62384 YDR522C | 0.1111 | 6 | 0.2000 | NO |
| | LOUVAIN | **YKR084C** YNL001W | 0.0833 | 2 | 1.0000 | NO |
| | EPCA | **YPR080W** YHR111W **YAL003W** YLR024C YPL236C **YPL048W** Q62384 YDR522C | 0.1875 | 8 | 0.2500 | NO |
| | EPD2 | **YPR080W** YHR111W **YAL003W** YLR024C YPL236C **YPL048W** Q62384 YDR522C | 0.1875 | 8 | 0.2500 | NO |
| | DIVANC | **YPR080W YAL003W** YLR024C YPL236C **YPL048W** Q62384 YDR522C | 0.2143 | 7 | 0.2857 | YES |

**Table C7.** The details about eEF-1 complex detected by overlapping algorithms.

| Benchmark/ Algorithms | Genes | Score | Size | Density | Success |
| --- | --- | --- | --- | --- | --- |
| **eEF-1 complex** | **YPL048W YKL081W YPR080W YBR118W YKR084C YAL003W** | - | 6 | 0.2667 | - |
| COACH | **YPR080W** YGL190C YLR249W **YKL081W** YGR240C | 0.1333 | 5 | 0.7000 | NO |
| ClusterOne | **YKR084C** YNL001W | 0.0833 | 2 | 1.0000 | NO |
| LinkComm | - | - | - | - | - |
| OSLOM2 | YJR035W YGL105W YML064C YMR059W YBR055C YJL138C **YKL081W** YLR442C YGR155W YMR226C YBL045C YKL035W YNR001C YDR023W YJL026W YMR318C YKL067W YFL045C YLR304C YKL085W YGL245W YJR104C YPL037C YKL095W YDR085C YPR191W YCL035C YBR025C YBL015W YFL022C YHR019C **YAL003W** YIR033W YNL192W YMR266W YLR060W YNR008W YIL037C | 0.0175 | 38 | 0.1209 | NO |
| EPD2' | **YPR080W** YHR111W **YAL003W** YLR024C YPL236C **YPL048W** Q62384 YDR522C | 0.1875 | 8 | 0.2500 | NO |
| DIVANC' | **YPR080W YAL003W** YLR024C YPL236C **YPL048W** Q62384 YDR522C | 0.2143 | 7 | 0.2857 | YES |

Table C8. The details about MIPS (116th) complex detected by non-overlapping algorithms.

| Benchmark/ Algorithms | Genes | Score | Size | Density | Success |
|---|---|---|---|---|---|
| **MIPS (116th) complex** | **YAL033W YHR062C YBR257W YBR167C YNL282W YNL221C SNRNA_NME1 YBL018C YGR030C** | - | 9 | 0.1944 | - |
| MCODE | - | - | - | - | - |
| INFOMAP | YDR226W YJR115W **YBL018C YBR167C** YOR176W **YAL033W YHR062C** YDR177W **YBR257W YGR030C** | 0.4000 | 10 | 0.2000 | YES |
| MCL | YJR115W **YBL018C YBR167C** YOR176W **YAL033W YHR062C YBR257W YGR030C** | 0.5000 | 8 | 0.2500 | YES |
| LOUVAIN | **YBL018C YBR167C YHR062C YBR257W YGR030C** | 0.5556 | 5 | 0.4000 | YES |
| EPCA | **YBL018C YBR167C YNL282W YAL033W YHR062C YBR257W YGR030C** | 0.7778 | 7 | 0.2857 | YES |
| EPD2 | **YBL018C YBR167C YNL282W YAL033W YHR062C YBR257W YGR030C** | 0.7778 | 7 | 0.2857 | YES |
| DIVANC | **YBL018C YBR167C YNL282W YAL033W YHR062C YBR257W YGR030C** | 0.7778 | 7 | 0.2857 | YES |

Table C9. The details about MIPS (116th) complex detected by overlapping algorithms.

| Benchmark/ Algorithms | Genes | Score | Size | Density | Success |
|---|---|---|---|---|---|
| **MIPS (116th) complex** | **YAL033W YHR062C YBR257W YBR167C YNL282W YNL221C SNRNA_NME1 YBL018C YGR030C** | - | 9 | 0.1944 | - |
| COACH | - | - | - | - | - |
| ClusterOne | **YAL033W** YJR115W | 0.0556 | 2 | 1.0000 | NO |
| LinkComm | - | - | - | - | - |
| OSLOM2 | YML100W YMR139W YOR303W YNL036W YER022W YMR134W YER138C YIL022W YHR152W YNL073W YIL063C YHL011C YNL281W YDR532C YHL045W YDR207C YBR003W YLR133W YML099C YDL001W YOL061W YMR079W YER132C YMR043W YDL091C YGL027C **YBL018C** | 0.0578 | 123 | 0.0184 | NO |

| | | | | | | |
|---|---|---|---|---|---|---|
| | YMR087W YNL210W YJR094C YOL106W YGR243W YML022W YHL044W YIL141W YJR148W YMR042W YMR293C YLL020C YOR187W YOL042W YNR024W YNL159C YML055W YPL098C YNR071C YMR039C YKL063C YJR126C **YNL282W** YER019W YLL065W YDL202W YBL065W YBR047W YEL012W YFR038W YGL261C YHL005C YHR116W YIL078W YJL142C YKL051W YKL141W YKR093W YLR331C YML082W YMR114C YMR118C YMR244W YNL211C YNL235C YOL022C YOR072W YOR143C YOR334W YOR336W YOR388C YPL053C YER160C YDR173C **YBR167C** YBL025W YDR170W-A YPL158C YJR099W YML014W YKL194C YLL027W YER131W YGL189C YLR435W YOR176W YER099C YKL181W YBL068W YPL196W YER163C YCR105W YOL081W YJR027W YJR028W YGL214W **YNL221C** YML107C YLR132C YDR441C YGL087C YKR035C YGL021W **YAL033W YHR062C** YOR090C YGR283C YPR065W YLR339C YFL044C YCR047C YNR002C YPL134C **YBR257W YGR030C** YOR130C | | | | | |
| EPD2' | **YBL018C YBR167C YNL282W YAL033W YHR062C YBR257W YGR030C** | 0.7778 | 7 | 0.2857 | YES |
| DIVANC' | **YBL018C YBR167C YNL282W YAL033W YHR062C YBR257W YGR030C** | 0.7778 | 7 | 0.2857 | YES |

**Table C10.** The benchmark and corresponding candidate metadata groups detected by DIVANC but cannot by EPD2 on *Hsa*HPRD.

| Candidate metadata groups | Size | Score | Benchmark | | Benchmark genes |
|---|---|---|---|---|---|

| | | | | |
|---|---|---|---|---|
| RAB33B **RABEP1 RABGEF1** HUNK | 4 | 0.2500 | RAB21-RABGEF1-RABEP1 complex | RAB21 **RABGEF1 RABEP1** HLA-B |
| SNTG1 **DMD** SNTG2 **PGM5** | 4 | 0.2500 | GO term golden standard(64th) | TNS1 **DMD** VCL **PGM5** |
| NUP214 **NXF1 TNPO1** NUP153 NXF5 | 5 | 0.2667 | NUP98-NXF1-TNPO1 complex | NUP98 **NXF1 TNPO1** |
| **CDK9 CCNT2** HEXIM1 ZMYM6 | 4 | 0.5000 | P-TEFb.2 complex | **CCNT2 CDK9** |
| **KPNA1 RAG1** ANP32A ELAVL1 | 4 | 0.3333 | KPNA1-RAG1-RAG2 complex | **KPNA1 RAG1** RAG2 |
| **MYCBP** PTCHD2 NP_060541.3 **MYCBPAP** C3orf15 | 5 | 0.2000 | C10orf10-MYCBP-MYCBPAP-VHL complex | **MYCBP** C10orf10 VHL **MYCBPAP** |
| **SRI GCA** | 2 | 1.0000 | Grancalcin-sorcin complex | **SRI GCA** |
| **DNM1L FIS1** | 2 | 1.0000 | DLP1-hFIS1 complex | **DNM1L FIS1** |

**Table C11.** The benchmark and corresponding candidate metadata groups detected by DIVANC but cannot by EPD2 on *Sce*DIP.

| Candidate metadata groups | Size | Score | Benchmark | Benchmark genes |
|---|---|---|---|---|
| YDL185W **YJR033C YDR202C** YJR062C | 4 | 0.3333 | RAVE complex | **YJR033C YDR202C** YDR328C |
| YJL130C **YGR258C YER041W** | 3 | 0.2667 | GO:0000014 | YHR164C **YGR258C** YGL175C TRM2 **YER041W** |
| **YGR092W** **YAL021C** **YFL028C** | 3 | 0.2308 | CCR4 complex | YPR072W **YGR092W** YCR093W YDL165W YNR052C YIL106W YDL160C **YFL028C** YER068W **YAL021C** YJR122W YKR036C YIL038C |
| YCR034W **YLL006W** **YOL009C** YOR112W | 4 | 0.2000 | ERMES complex | YAL010C **YLL006W** **YOL009C** YAL048C YGL219C |

**Table C12.** The candidate metadata groups detected by DIVANC and DIVANC', while the latter matching their own benchmarks perfectly.

| Benchmark | Genes by DIVANC | Score of DIVANC | Genes by DIVANC' | Score of DIVANC' |
|---|---|---|---|---|
| Ribose phosphate diphosphokinase complex | YHL011C YKL181W YBL068W YER099C | 0.8000 | YKL181W YER099C YHL011C YBL068W YOL061W | 1.000 |
| C1orf63-CLK2-CLK3-YWHAG complex | CLK2 CLK3 C1orf63 | 0.7500 | CLK2 CLK3 C1orf63 YWHAG | 1.000 |
| ITGA9-ITGB1-SPP1 complex | SPP1 ITGA9 | 0.6667 | SPP1 ITGA9 ITGB1 | 1.000 |
| AXIN-MEKK4-CCD1 complex | MAP3K4 DIXDC1 | 0.6667 | AXIN1 DIXDC1 MAP3K4 | 1.000 |
| CCBL2-HBXIP-RABIF-UTP14A complex | RABIF HBXIP CCBL2 | 0.7500 | RABIF HBXIP CCBL2 UTP14A | 1.000 |
| Tumor necrosis family (TNF)/TNF receptor (TNFR) complex | TNFRSF17 TNFSF13B TNFRSF13C TNFRSF13B | 0.8000 | TNFRSF17 TNFSF13B TNFRSF13C TNFRSF13B TNFSF13 | 1.000 |

**References**

[1]     Fortunato S 2010 Community detection in graphs *Phys. Rep.* **486** 75–174

[2]     Enright A J, Van Dongen S and Ouzounis C A 2002 An efficient algorithm for large-scale detection of protein families *Nucleic Acids Res.* **30** 1575–84

[3]    Rosvall M and Bergstrom C T 2008 Maps of random walks on complex networks reveal community structure *Proc. Natl. Acad. Sci.* **105** 1118–23

[4]    Bader G D and Hogue C W V 2003 An automated method for finding molecular complexes in large protein interaction networks *BMC Bioinformatics* **4** 2

[5]    Nepusz T, Yu H and Paccanaro A 2012 Detecting overlapping protein complexes in protein-protein interaction networks *Nat. Methods* **9** 471–2

[6]    Ahn Y-Y, Bagrow J P and Lehmann S 2010 Link communities reveal multiscale complexity in networks *Nature* **466** 761–4

[7]    Blondel V D, Guillaume J-L, Lambiotte R and Lefebvre E 2008 Fast unfolding of communities in large networks *J. Stat. Mech. Theory Exp.* **2008** P10008

[8]    Lancichinetti A, Radicchi F, Ramasco J J and Fortunato S 2011 Finding Statistically Significant Communities in Networks *PLoS ONE* **6** e18961

[9]    Zhang J, Zhang K, Xu X, Chi K T and Small M 2009 Seeding the kernels in graphs: Toward multi-resolution community analysis *New J. Phys.* **11** 113003

[10]   Deritei D, Lázár Z I, Papp I, Járai-Szabó F, Sumi R, Varga L, Regan E R and Ercsey-Ravasz M 2014 Community detection by graph Voronoi diagrams *New J. Phys.* **16** 063007

[11]   Jia S, Gao L, Gao Y, Nastos J, Wang Y, Zhang X and Wang H 2015 Defining and identifying cograph communities in complex networks *New J. Phys.* **17** 013044

[12]   Radicchi F 2014 A paradox in community detection *EPL Europhys. Lett.* **106** 38001

[13]   Yang J and Leskovec J 2013 Overlapping community detection at scale: a nonnegative matrix factorization approach *Proceedings of the sixth ACM international conference on Web search and data mining* (ACM) pp 587–96

[14]   Yang J and Leskovec J 2013 Defining and evaluating network communities based on ground-truth *Knowl. Inf. Syst.* **42** 181–213

[15]   Hric D, Darst R K and Fortunato S 2014 Community detection in networks: Structural communities versus ground truth *Phys. Rev. E* **90** 062805

[16]   Yang J and Leskovec J 2014 Structure and overlaps of ground-truth communities in networks *ACM Trans. Intell. Syst. Technol. TIST* **5** 26

[17]   Balasubramanyan R and Cohen W W 2011 Block-LDA: Jointly modeling entity-annotated text and entity-entity links. *SDM* vol 11 (SIAM) pp 450–61

[18]   Ruan Y, Fuhry D and Parthasarathy S 2013 Efficient community detection in large networks using content and links *Proceedings of the 22nd international conference on world wide web* (International World Wide Web Conferences Steering Committee) pp 1089–98

[19] Pool S, Bonchi F and Leeuwen M van 2014 Description-driven community detection *ACM Trans. Intell. Syst. Technol. TIST* **5** 28

[20] Balasov M, Huijbregts R P and Chesnokov I 2009 Functional analysis of an Orc6 mutant in Drosophila *Proc. Natl. Acad. Sci.* **106** 10672–7

[21] Santt O, Pfirrmann T, Braun B, Juretschke J, Kimmig P, Scheel H, Hofmann K, Thumm M and Wolf D H 2008 The yeast GID complex, a novel ubiquitin ligase (E3) involved in the regulation of carbohydrate metabolism *Mol. Biol. Cell* **19** 3323–33

[22] Wolf D M and Arkin A P 2003 Motifs, modules and games in bacteria *Curr. Opin. Microbiol.* **6** 125–34

[23] Yeger-Lotem E, Sattath S, Kashtan N, Itzkovitz S, Milo R, Pinter R Y, Alon U and Margalit H 2004 Network motifs in integrated cellular networks of transcription–regulation and protein–protein interaction *Proc. Natl. Acad. Sci. U. S. A.* **101** 5934–9

[24] Albert I and Albert R 2004 Conserved network motifs allow protein–protein interaction prediction *Bioinformatics* **20** 3346–52

[25] Yook S-H, Oltvai Z N and Barabási A-L 2004 Functional and topological characterization of protein interaction networks *Proteomics* **4** 928–42

[26] Salwinski L, Miller C S, Smith A J, Pettit F K, Bowie J U and Eisenberg D 2004 The database of interacting proteins: 2004 update *Nucleic Acids Res.* **32** D449–51

[27] Prasad T K, Goel R, Kandasamy K, Keerthikumar S, Kumar S, Mathivanan S, Telikicherla D, Raju R, Shafreen B and Venugopal A 2009 Human protein reference database—2009 update *Nucleic Acids Res.* **37** D767–72

[28] Lancichinetti A, Fortunato S and Radicchi F 2008 Benchmark graphs for testing community detection algorithms *Phys. Rev. E* **78** 046110

[29] Lancichinetti A and Fortunato S 2009 Benchmarks for testing community detection algorithms on directed and weighted graphs with overlapping communities *Phys. Rev. E* **80** 016118

[30] Girvan M and Newman M E J 2002 Community structure in social and biological networks *Proc. Natl. Acad. Sci.* **99** 7821–6

[31] Evans T S 2010 Clique graphs and overlapping communities *J. Stat. Mech. Theory Exp.* **2010** P12037

[32] Mewes H-W, Amid C, Arnold R, Frishman D, Güldener U, Mannhaupt G, Münsterkötter M, Pagel P, Strack N and Stümpflen V 2004 MIPS: analysis and annotation of proteins from whole genomes *Nucleic Acids Res.* **32** D41–4

[33] Hong E L, Balakrishnan R, Dong Q, Christie K R, Park J, Binkley G, Costanzo M C, Dwight S S,

Engel S R and Fisk D G 2008 Gene Ontology annotations at SGD: new data sources and annotation methods *Nucleic Acids Res.* **36** D577–81

[34] Shih Y-K and Parthasarathy S 2012 Identifying functional modules in interaction networks through overlapping Markov clustering *Bioinformatics* **28** i473–9

[35] Wang Y and Qian X 2014 Functional module identification in protein interaction networks by interaction patterns *Bioinformatics* **30** 81–93

[36] Kikugawa S, Nishikata K, Murakami K, Sato Y, Suzuki M, Altaf-Ul-Amin M, Kanaya S and Imanishi T 2012 PCDq: human protein complex database with quality index which summarizes different levels of evidences of protein complexes predicted from H-Invitational protein-protein interactions integrative dataset *BMC Syst. Biol.* **6** S7

[37] Ruepp A, Brauner B, Dunger-Kaltenbach I, Frishman G, Montrone C, Stransky M, Waegele B, Schmidt T, Doudieu O N, Stümpflen V and Mewes H W 2008 CORUM: the comprehensive resource of mammalian protein complexes *Nucleic Acids Res.* **36** D646–50

[38] Serrour B, Arenas A and Gómez S 2011 Detecting communities of triangles in complex networks using spectral optimization *Comput. Commun.* **34** 629–34

[39] Corneil D G, Lerchs H and Burlingham L S 1981 Complement reducible graphs *Discrete Appl. Math.* **3** 163–74

[40] Jia S, Gao L, Gao Y and Wang H 2014 Anti-triangle centrality-based community detection in complex networks *IET Syst. Biol.* **8** 116–25

[41] Radicchi F, Castellano C, Cecconi F, Loreto V and Parisi D 2004 Defining and identifying communities in networks *Proc. Natl. Acad. Sci. U. S. A.* **101** 2658–63

[42] Palla G, Derényi I, Farkas I and Vicsek T 2005 Uncovering the overlapping community structure of complex networks in nature and society *Nature* **435** 814–8

[43] Wu M, Li X, Kwoh C-K and Ng S-K 2009 A core-attachment based method to detect protein complexes in PPI networks *BMC Bioinformatics* **10** 169

[44] Csardi G and Nepusz T 2006 The igraph software package for complex network research *InterJournal Complex Syst.* 1965

[45] Kalinka A T and Tomancak P 2011 linkcomm: an R package for the generation, visualization, and analysis of link communities in networks of arbitrary size and type *Bioinformatics* **27**

[46] Li X, Wu M, Kwoh C-K and Ng S-K 2010 Computational approaches for detecting protein complexes from protein interaction networks: a survey *BMC Genomics* **11** S3

[47] Yamasaki C, Murakami K, Takeda J, Sato Y, Noda A, Sakate R, Habara T, Nakaoka H, Todokoro F, Matsuya A, Imanishi T and Gojobori T 2010 H-InvDB in 2009: extended database and data mining resources for human genes and transcripts *Nucleic Acids Res.* **38** D626–32

[48]   McDaid A F, Greene D and Hurley N 2011 Normalized mutual information to evaluate overlapping community finding algorithms *ArXiv Prepr. ArXiv11102515*

[49]   Lancichinetti A, Fortunato S and Kertész J 2009 Detecting the overlapping and hierarchical community structure in complex networks *New J. Phys.* **11** 033015

[50]   Cho Y-R, Hwang W, Ramanathan M and Zhang A 2007 Semantic integration to identify overlapping functional modules in protein interaction networks *BMC Bioinformatics* **8** 265